\documentclass{aastex61}

\usepackage{graphicx}	
\usepackage{amssymb}
\usepackage{amsmath}
\usepackage{natbib}
\usepackage{color}
\usepackage{url}
\usepackage{placeins}
\usepackage{float}
\usepackage{color}
\usepackage{rotating}

\newcommand\aastex{AAS\TeX}

\newcommand{\ie}{{\it i.e.\,}}
\newcommand{\eg}{{\it e.g.}}
\newcommand{\insitu}{{\it in situ~}}
\newcommand{\kmps}{{$\mathrm{~km~s}^{-1}$}}
\newcommand{\Dunit}{{~$\mathrm{cm^{2}~s^{-1}}$\,}}
\newcommand{\gammaunit}{{$\times10^{-7} ~\mathrm{km}^{-1}$\,}}
\newcommand{\Rsun}{{$~\mathrm{R}_{\odot}$}}

\submitjournal{ApJ}

\shorttitle{\aastex\ Forbush Decrease Model (ForbMod)}
\shortauthors{Dumbovi\'{c} et al.}

\begin{document}

\title{An analytical diffusion-expansion model for Forbush decreases caused by flux ropes}

\correspondingauthor{Mateja Dumbovi\'{c}}
\email{mateja.dumbovic@uni-graz.at}

\author[0000-0002-8680-8267]{Mateja Dumbovi\'{c}}
\affil{Institute of Physics, University of Graz, Universit\"atsplatz 5, A-8010 Graz, Austria}

\author[0000-0003-0960-5658]{Bernd Heber}
\affiliation{Department of Extraterrestrial Physics, Christian-Albrechts University in Kiel, Liebnitzstrasse 11, 24098, Kiel, Germany}

\author[0000-0002-0248-4681]{Bojan Vr\v{s}nak}
\affiliation{Hvar Observatory, Faculty of Geodesy, University of Zagreb, Ka\v{c}i\'{c}eva 26, HR-10000, Zagreb, Croatia}

\author[0000-0003-4867-7558]{Manuela Temmer}
\affil{Institute of Physics, University of Graz, Universit\"atsplatz 5, A-8010 Graz, Austria}

\author[0000-0002-6873-5076]{Anamarija Kirin}
\affil{Karlovac University of Applied Sciences, Karlovac, Croatia}

\begin{abstract}

We present an analytical diffusion--expansion Forbush decrease (FD) model \textit{ForbMod} which is based on the widely used approach of the initially empty, closed magnetic structure (\ie flux rope) which fills up slowly with particles by perpendicular diffusion. The model is restricted to explain only the depression caused by the magnetic structure of the interplanetary coronal mass ejection (ICME). We use remote CME observations and a 3D reconstruction method (the Graduated Cylindrical Shell method) to constrain initial boundary conditions of the FD model and take into account CME evolutionary properties by incorporating flux rope expansion. Several flux rope expansion modes are considered, which can lead to different FD characteristics. In general, the model is qualitatively in agreement with observations, whereas quantitative agreement depends on the diffusion coefficient and the expansion properties (interplay of the diffusion and the expansion). A case study was performed to explain the FD observed 2014 May 30. The observed FD was fitted quite well by \textit{ForbMod} for all expansion modes using only the diffusion coefficient as a free parameter, where the diffusion parameter was found to correspond to expected range of values. Our study shows that in general the model is able to explain the global properties of FD caused by FR and can thus be used to help understand the underlying physics in case studies.

\end{abstract}


\section{Introduction}
\label{intro}

Forbush decreases (FDs) are depressions in the galactic cosmic ray (GCR) count-rate observed around the passage of solar wind transients such as interplanetary coronal mass ejections (ICMEs), stream interaction regions (SIRs), and interplanetary shocks \citep{forbush37,lockwood71,cane00,belov09}. ICMEs produce the strongest FDs, which often show a two-step profile, one associated with the shock/sheath region and the other with ICME magnetic structure \citep{barnden73,cane00}. The two regions were found to be roughly equally effective in GCR modulation and moreover the modulation of the ICME magnetic structure is found to be particularly effective for magnetic clouds \citep{richardson11a}. Magnetic clouds are a subset of ICMEs characterised by low proton temperature, low proton beta parameter and most notably by smoothly rotating magnetic field, indicative of a flux rope magnetic structure \citep{burlaga81,zurbuchen06,rouillard11}, where a flux rope (FR) is a cylindrical plasma structure with magnetic field lines helically winding around the central axis \citep{lepping90}.

Generally, solar activity modulates GCRs, which can be described by a Fokker-Planck transport equation of particle random walk in the frame of reference of the small-scale magnetic irregularities \citep{parker65}. Based on this so-called \textit{Parker equation} the four physical mechanisms governing the GCR modulation are diffusion, drifts, convection and energy loss \citep[see also \eg][and references therein]{jokipii71,potgieter13}. FDs can be considered as the disturbances in the GCR distribution caused by local variations in one or more transport parameters. However, it is important to separately model FDs corresponding to different solar wind transients. The shock/sheath region is the region of disturbed conditions and highly fluctuating magnetic field ahead of the ICME and is magnetically connected to the ambient interplanetary space \citep[\eg][and references therein]{kilpua17}. Therefore, shock-associated FD can be modelled as \eg\, propagating diffusive barrier \citep{wibberenz98} or numerically, by solving the Parker equation \citep[\eg][]{leroux91,wawrzynczak10,alania13}. The ICME magnetic structure is not magnetically connected to the ambient plasma and is characterised by smooth magnetic field \citep[\eg][and references therein]{kilpua17}. Recently, \citet{jordan11} revisited the two-step nature of FDs attributing the variety of FD profiles to small-scale interplanetary magnetic field structures, but nevertheless since the two regions are characterised by different properties it is reasonable to regard them as two different global structures with respect to FD modelling, especially given that they are not always encountered together \citep{richardson11a}.

It was first proposed by \citet{morrison56} that FDs might be clouds of closed magnetised plasma, where they are initially empty of particles and slowly fill as they propagate through the interplanetary space. This approach has been since utilised in several studies to explain FR-associated FDs, where the particles are allowed to enter via ``cross-field" transport, \ie perpendicular diffusion \citep{cane95,munakata06,quenby08,subramanian09} and/or drifts \citep{krittinatham09} and gyration \citep{kubo10}. Strictly speaking, the transport across the FR boundary cannot easily be explained by the simple perpendicular (sub)diffusion, because the field lines on opposite sides of the FR boundary are not connected \citep{jokipii66,ruffolo08,krittinatham09}. However, the ``cross-field" transport also refers to another mechanism which considers a situation where the particle motion is not bound to its original guiding field line, resulting in diffusion at late times \citep{ruffolo08,krittinatham09}. Furthermore, a change of the magnetic connectivity of the FR can occur due to reconnection allowing magnetic fluctuations (and therefore particles) to diffuse inside FR \citep{masias-meza16}. \citet{krittinatham09} have shown that particles can enter \textit{via} gradient and curvature drifts, however, according to their simulations drift orbits in the FR were spatially restricted to the areas close to the FR boundary and a large fraction exits the FR after several minutes. Similar results were obtained by \citet{kubo10} from an analytical model based on gyration. The conclusion in both studies was that diffusion is needed to explain the transport of particles into the interior of the FR. The diffusion approach is generally considered valid when the gyroradius is smaller than the size of the FR \citep[\eg][]{subramanian09}, and \citet{blanco13b} showed that for FRs detected by \textit{Helios} spacecraft typical rigidity of a GCR with gyroradius corresponding to the size of FR is of the order of $\sim100$ GV. Therefore, the diffusion approach should be valid to explain FR associated FDs detected in a variety of instruments corresponding to different mean energies: muon telescopes at $\sim50$ GV \citep[\eg][]{kozai16}, neutron monitors at $\sim10$ GV \citep[\eg][]{clem00}, and spacecraft such as the \textit{Electron Proton Helium Instrument} \citep[EPHIN,][]{muller-mellin95} onboard \textit{Solar and Heliospheric Observatory} \citep[\textit{SOHO},][]{domingo95} at $\sim1$ GV \citep[\eg][]{kuhl15}. The models based on perpendicular diffusion \citep{cane95,munakata06,quenby08,subramanian09} are in a good qualitative agreement with the observation, reflecting some of the observational characteristics of FDs caused by ejecta such as symmetric shape constrained to the spatial extent of the flux rope and relation of the FD amplitude to the magnetic field strength, the FR size and ICME transit time \citep{cane93,belov09,dumbovic11,dumbovic12b,blanco13b,badruddin16,masias-meza16}, although there are evidences that simple diffusion FD models cannot fully explain observations \citep{richardson11a}.

In addition to the particle diffusion, another important contribution comes from the expansion of the FR. It was first proposed by \citet{laster62} and \citet{singer62} that expansion is needed to explain the observational properties of FDs. More recently the diffusion--expansion approach was applied by \citet{munakata06} and \citet{kuwabara09} in a numerical model best-fitted to the measurements taken by the muon network at Earth (rigidities 50-100 GV) and was utilised to determine the orientation of the interplanetary flux rope. The diffusion--expansion approach was also implemented by \citet{subramanian09} and \citet{arunbabu13} in an analytical model and was compared to GRAPES-3 muon telescope (rigidities 12-42 GV). These authors estimated the radial perpendicular diffusion coefficient to be of the order of $10^{21}$\Dunit, however, it should be noted that their studies are adjusted to large events associated with a shock and observed at high particle rigidities. At rigidities of about 1 GV \citet{cane95} estimated a diffusion coefficient of about $10^{19}$\Dunit utilising spacecraft measurements and less energetic ejecta. Both are roughly in agreement with values obtained from the typical empirical expression used in numerical models \citep[see Equations 23 and 24 in][]{potgieter13}. Therefore, these studies outline a rough constraint on the diffusion coefficient used to model FR-associated FD. Since both diffusion and expansion are processes which presumably start close to the Sun, FR \ie CME initial properties need to be considered in the model. This ejecta-only FD model primarily describes FDs associated with ICMEs without shock/sheath region, because shock/sheath region presumably introduces additional decrease \textit{via} different physical mechanisms, but could also be applied to explain the ejecta part of FD in two-step FDs. The aim of this study is to take these considerations into account in the diffusion--expansion model to improve our understanding on the cause, formation, and evolution of FDs.

\section{The basic diffusion model}
\label{model1}	

First we consider an analytical FD model based on the perpendicular diffusion of GCRs into the FR, similar to \citet{cane95}. In this basic diffusion model the FR is regarded as a closed magnetic structure, rooted at the Sun, which is initially empty of GCRs (Figure \ref{fig1}a). The FR is of cylindrical form (hereafter approximated by a long cylinder), it moves with constant velocity and as it moves, it does not vary in shape or size (Figure \ref{fig1}b). We note that the assumption of constant velocity is in general not valid, as it was shown that CMEs slower than the solar wind accelerate, whereas CMEs faster than the solar wind decelerate \citep[\eg][]{sheeley99,gopalswamy00}, which is attributed to the magnetohydrodynamical drag \citep[see \eg][and references therein]{vrsnak13,hess14,sachdeva15}. However, since the acceleration/deceleration depends on the difference between the CME and solar wind speed, for a substantial subset of CMEs (and especially the slower ones which do not drive shocks) constant velocity can be taken as a relatively fair approximation. We also note that the assumption of constant shape and size might not hold true, as it was shown that CMEs expand while propagating in the interplanetary space \citep[\eg][]{bothmer98,leitner07,janvier14} and that their shape might deform during propagation \citep[\eg][]{cargill94,liu06}. The deformation of shape is mainly kinematic effect, therefore, the approximation of the constant shape is closely related to the constant velocity approximation. On the other hand, CME expansion is an important feature that may drastically influence the interaction of CMEs and GCRs and will be addressed in the next section.

Furthermore, it is assumed that the GCR density outside the FR is constant and that the GCR can enter the interior of the FR only \textit{via} perpendicular, \ie radial diffusion. It has been shown that there is a radial gradient of GCRs of about $\sim3\%$/AU \citep[\eg][]{webber99,gieseler16,lawrence16}. Recently, \citet{marquardt18} used Helios E6 data to show that the gradient of the anomalous cosmic ray oxygen is increasing with decreasing distance to Sun. Therefore, the assumption of constant GCR density outside the FR is not entirely correct. However, for simplicity we assume that the change of the outside density of $\sim3\%$ throughout the evolution of the FR to 1 AU will not notably influence the filling rate of GCRs into the FR and therefore we keep the assumption that the GCR density outside the FR is constant (the estimation to justify this assumption is given at the end of Section \ref{model2}).

With the assumptions listed above the density of GCRs can be calculated based on the radial diffusion equation for a long cylinder given by \eg\, \citet{crank}:

\begin{equation}
 \frac{\partial U}{\partial t} = \frac{1}{r}\Bigg(\frac{\partial}{\partial r}\Big(rD_{\perp}\frac{\partial}{\partial r}\Big)\Bigg)\,,
\label{eq1}
\end{equation}

\noindent where $U=U(r,P,t)$ is the GCR phase space density, $D_{\perp}$ is the perpendicular (radial) diffusion coefficient related to GCR rigidity, $P$ (here-forth denoted simply as $D$), $t$ is time, and $r$ is the radial distance from the FR centre. For particles of specific rigidity $U=U(r,t)$ and the partial differential equation is solved using the method of separation of variables ($U(r,t)=T(t)R(r)$), under the assumption that the diffusion coefficient does not depend on $r$. The time dependence is then given by the expression \citep[see \eg][for details]{crank,butkov}:

\begin{equation}
 T(t) = \mathrm{e}^{-\lambda^2Dt}\,,
\label{eq2}
\end{equation}

\noindent where $\lambda$ is a constant determined by the initial and boundary conditions. It can be shown that the equation for the radial dependance can be written in a form:

\begin{equation}
 r^2R(r)''+rR(r)'+\lambda^2r^2R(r)=0\,,
\label{eq3}
\end{equation}

\noindent which is the Bessel's equation of the order 0. The solution of Equation \ref{eq3} can be generally written as a linear combination of the Bessel and Neumann functions, $J_0(\lambda r)$ and $N_0(\lambda r)$, respectively. However, the solution must be finite in the center of the FR, \ie $R(\lambda r)$ finite at $r=0$, and thus only $J_0(\lambda r)$ is admissible as a solution \citep[$N_0(\lambda r)$ is not finite in $r=0$, see \eg][for details]{butkov}. Therefore, the general solution of Equation \ref{eq1} is:

\begin{equation}
 U(r,t) = CJ_0(\lambda r)\mathrm{e}^{-\lambda^2Dt}\,,
\label{eq4}
\end{equation}

\noindent where $C$ and $\lambda$ are constants determined by the initial and boundary conditions, which can be written in the form:

\begin{equation}
	U(r,t) =
	\begin{cases}
	0, & 0<r<a, t=0 \\
	U_0, & r=a, t\ge0
	\end{cases}
\label{eq5}
\end{equation}

\noindent $a$ is the radius of the FR and $U_0$ is the GCR phase space density at its surface. With these initial and boundary conditions, the solution for the particle density inside the flux rope can be written  \citep[see Equation 5.22 in][]{crank}:

\begin{equation}
 U(r,t) = U_0\Bigg(1-\frac{2}{a}\sum_{n=1}^{\infty}\frac{J_0(\lambda_nr)}{\lambda_nJ_1(\lambda_na)}\mathrm{e}^{-D\lambda_n^2t}\Bigg)\,,
\label{eq6}
\end{equation}

\noindent where $J_0$ and $J_1$ are Bessel functions (of the first kind) of the order 0 and 1, respectively, and $\lambda_n$ are defined by the positive roots of $J_0(\lambda_na)=0$ ($\lambda_n = \frac{\alpha_n}{a}$, $\alpha_n$ are positive roots of $J_0$), which are tabulated in tables of Bessel functions. Bessel functions $J_0$ have oscillatory character whereas the exponential function rapidly decreases with $\alpha_n^2$, therefore the solution can be approximated as:

\begin{equation}
 U(r,t) = U_0\Bigg(1-C\frac{2}{\alpha_1}\frac{J_0(\alpha_1\frac{r}{a})}{J_1(\alpha_1)}\mathrm{e}^{-D(\frac{\alpha_1}{a})^2t}\Bigg)\,,
\label{eq7}
\end{equation}

\noindent where the constant $C$ depends on the initial and boundary conditions defined in Equation \ref{eq5}. The final solution is then given by the expression:

\begin{equation}
 U(r,t) = U_0\Bigg(1-J_0(\alpha_1\frac{r}{a})\mathrm{e}^{-D(\frac{\alpha_1}{a})^2t}\Bigg)\,.
\label{eq8}
\end{equation}

\noindent The corresponding Forbush decrease can be represented by $A(\%)=(U(r,t)/U_0-1)\cdot100\%$, where the radial dependence for a given diffusion time depicts its shape and magnitude at a certain location in the heliosphere.

\section{The diffusion--expansion model}
\label{model2}

\begin{figure*}
\gridline{\fig{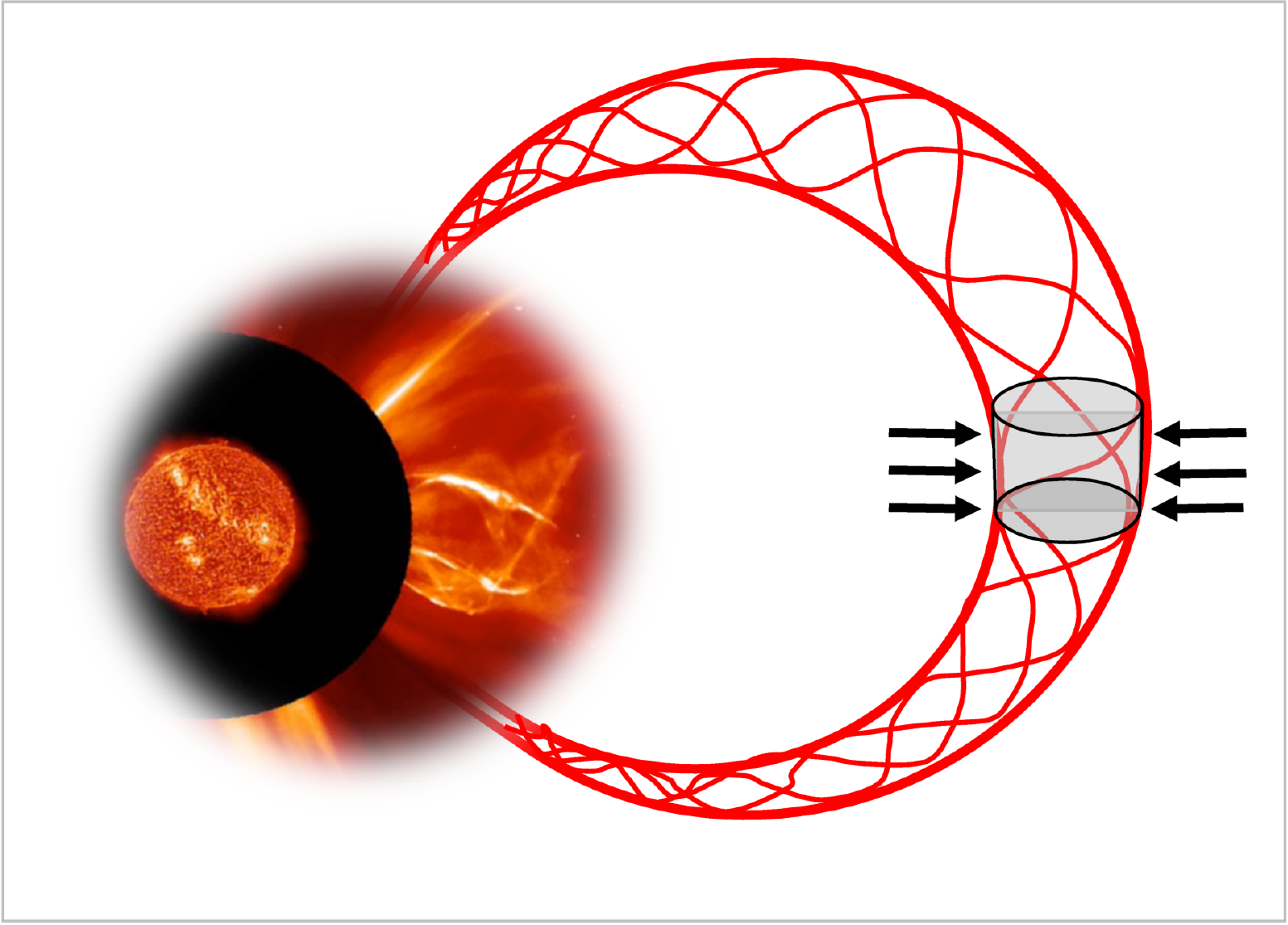}{0.45\textwidth}{(a)}
          \fig{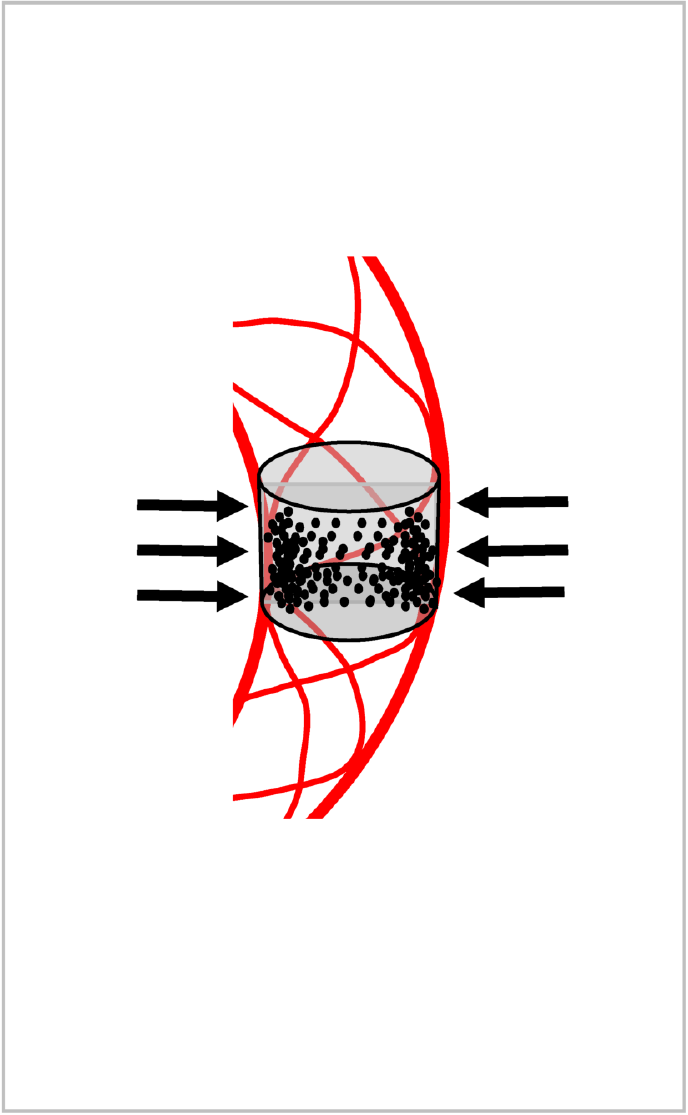}{0.2\textwidth}{(b)}
           \fig{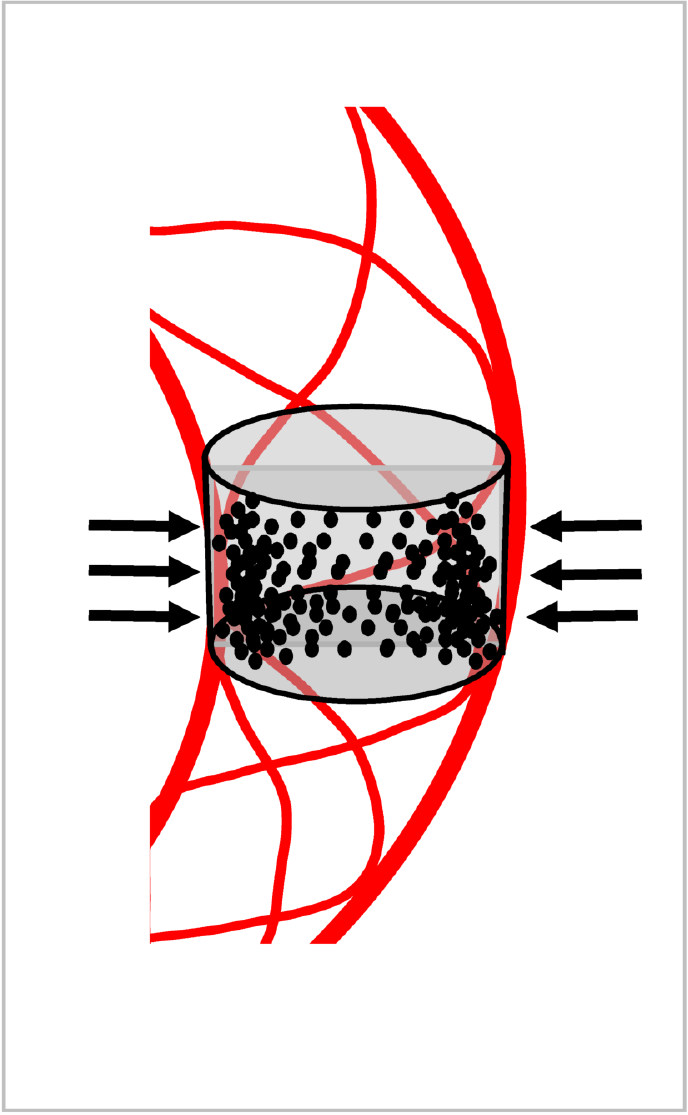}{0.2\textwidth}{(c)}
          }
\caption{a) A sketch of the initial FR for both diffusion-only and diffusion-expansion model: FR is a closed magnetic structure locally of the cylindrical form, rooted at the Sun and initially empty of GCRs; b) A sketch of the diffusion-only model after time $t$: FR does not vary in shape or size; c) A sketch of the diffusion-expansion model after time $t$: FR expands self-similarly. In both cases particles enter the FR by perpendicular diffusion. Image of the Sun is adapted from the remote EUV and coronagraphic observations by \textit{SOHO}.
\label{fig1}}
\end{figure*}

Now we introduce the expansion of the FR by allowing that the radius changes with time (Figure \ref{fig1}c). We note that we do not take into account the energy-loss effect (adiabatic cooling) due to FR expansion. In general, the energy-loss is expected to introduce additional modulation effects \citep[see \eg][and references therein]{lockwood71}, but the quantitative contribution is not trivial to estimate. The energy-loss term is proportional to $\partial U/\partial lnP$ and to some extent balances out the inward diffusion of particles \citep{munakata06}. Therefore, it is reasonable to expect that neglecting the GCR energy loss could result in a somewhat underestimated FD amplitudes.

Furthermore, we allow the diffusion coefficient to change with time as well, since it presumably depends on the magnetic field strength \citep{potgieter13}, which in general also changes in time \citep[\eg][]{demoulin08}. Consequently, the dependence on the radial distance and time are not independent and the method of separation of variables is no longer applicable to Equation \ref{eq1}. However, with the substitution $r(t)\rightarrow \hat{r}(t)/a(t)$ Equation \ref{eq1} can be rewritten in a form:

\begin{equation}
 \frac{\partial U}{\partial t} =\frac{D(t)}{a(t)^2} \frac{1}{\hat{r}}\Bigg(\frac{\partial}{\partial\hat{r}}\Big(\hat{r}\frac{\partial}{\partial\hat{r}}U\Big)\Bigg)\,,
\label{eq9}
\end{equation}

\noindent where $\hat{r}$ is the normalized radial distance, scaled to the radius of the flux rope ($\hat{r}=r(t)/a(t)$), whereas $D(t)$ and $a(t)$ are time-dependent diffusion coefficient and flux rope radius, respectively. With the assumption that the ratio $r(t)/a(t)$ does not change with time (which is discussed below), the method of separation of variables can be applied to Equation \ref{eq9}. The form of the radial solution remains the same, whereas the time-dependent solution can be written in the form:

\begin{equation}
 T(t) = \mathrm{e}^{-\lambda^2\int\frac{D(t)}{a(t)^2}dt}\,.
\label{eq10}
\end{equation}

\noindent Note that the initial and boundary conditions need to be rewritten accordingly:

\begin{equation}
	U(\hat{r},t) =
	\begin{cases}
	0, & 0<\hat{r}<1, t=0 \\
	U_0, & \hat{r}=1, t\ge0
	\end{cases}
\label{eq11}
\end{equation}

\noindent Following the same procedure as described in Section \ref{model1} the final solution can be written as:

\begin{equation}
 U(\hat{r},t) = U_0\Bigg(1-J_0(\alpha_1\hat{r})\mathrm{e}^{-\alpha_{1}^{2}f(t)}\Bigg)\,,
\label{eq12}
\end{equation}

\noindent where $U_0$ is the GCR phase space density at the FR surface, $J_0$ is a Bessel function (of the first kind) of the order 0, $\alpha_1$ is a first positive root of $J_0$ (tabulated in tables of Bessel functions), $\hat{r}$ is the radial distance scaled to the FR radius ($\hat{r}=r(t)/a(t)$), and $f(t)$ is a function of time that depends on the interplay between the diffusion of GCR into the FR and its expansion. Since $f(t)$ depends on the ratio $D(t)/a(t)^2$, where both $D(t)$ and $a(t)$ are generally not known and are still  subject of ongoing studies, it is a somewhat arbitrary function. However, there are some basic constraints on the function $f(t)$. Firstly, it must not diverge at $t=0$ and moreover, due to the initial condition, it must satisfy $f(t=0)=0$. Secondly, the behaviour of the function is constrained by observations. \citet{cane94} and \citet{blanco13b} found that the FD amplitude decreases with heliospheric distance, indicating that the GCR density within the FR increases with time. Therefore, we expect $f(t)$ to be a positive and monotonically increasing function. Further constraints on the function $f(t)$, \ie the solution given in Equation \ref{eq12}, can be achieved through comparison of the model results with observations, while a general behaviour is derived in the continuation of this Section based on empirical FR relations.

\begin{figure*}
\gridline{\fig{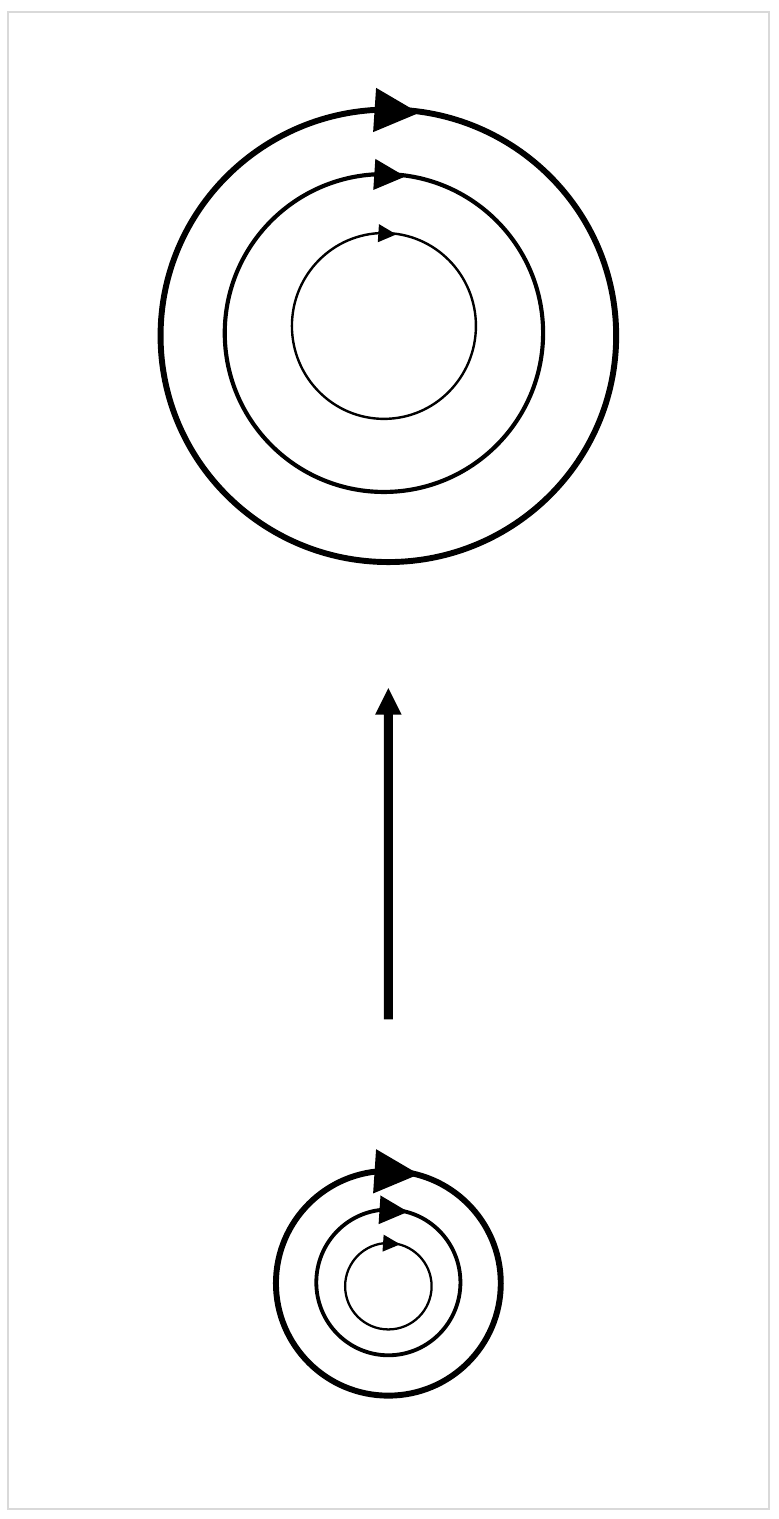}{0.3\textwidth}{(a)}
          \fig{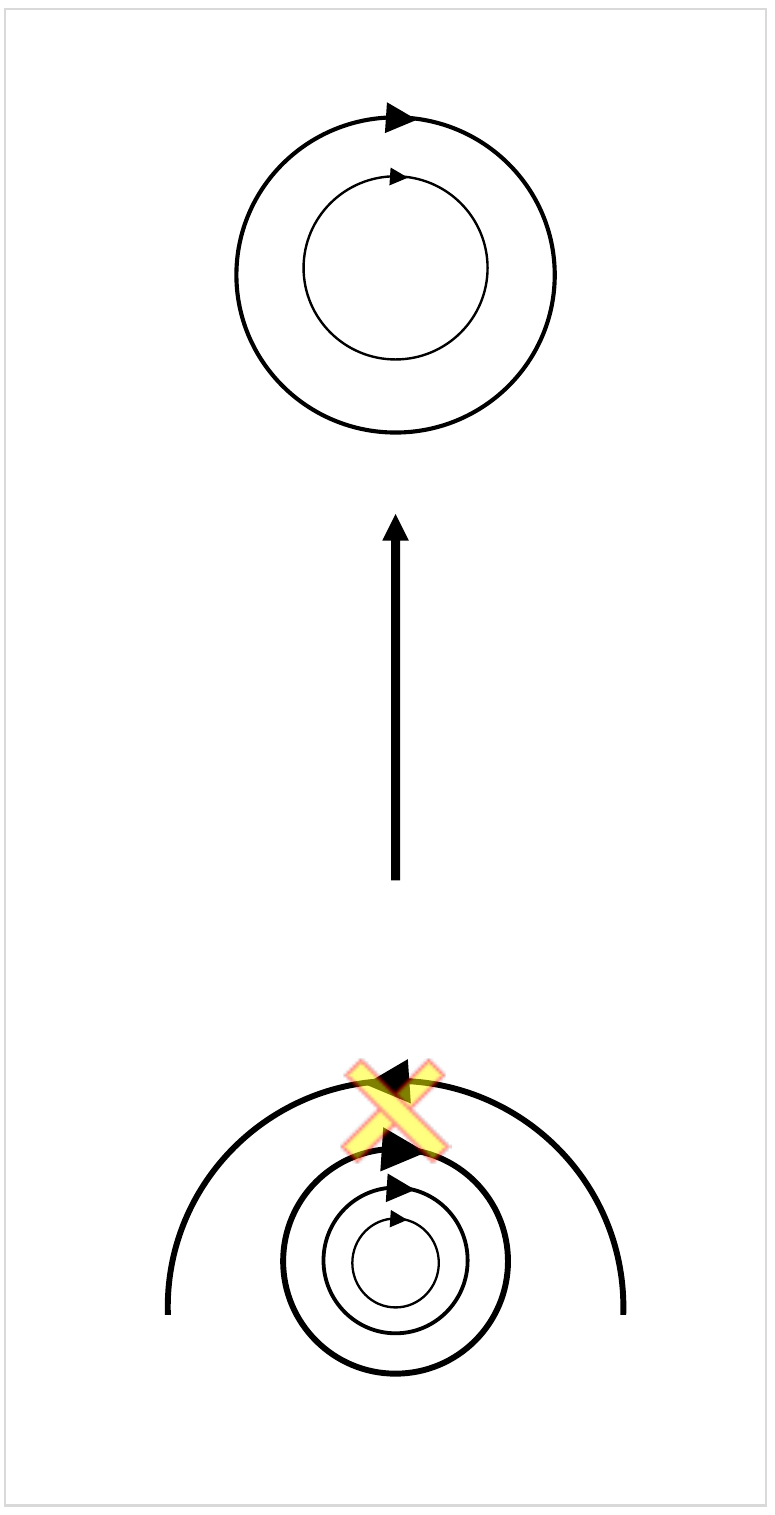}{0.3\textwidth}{(b)}
           \fig{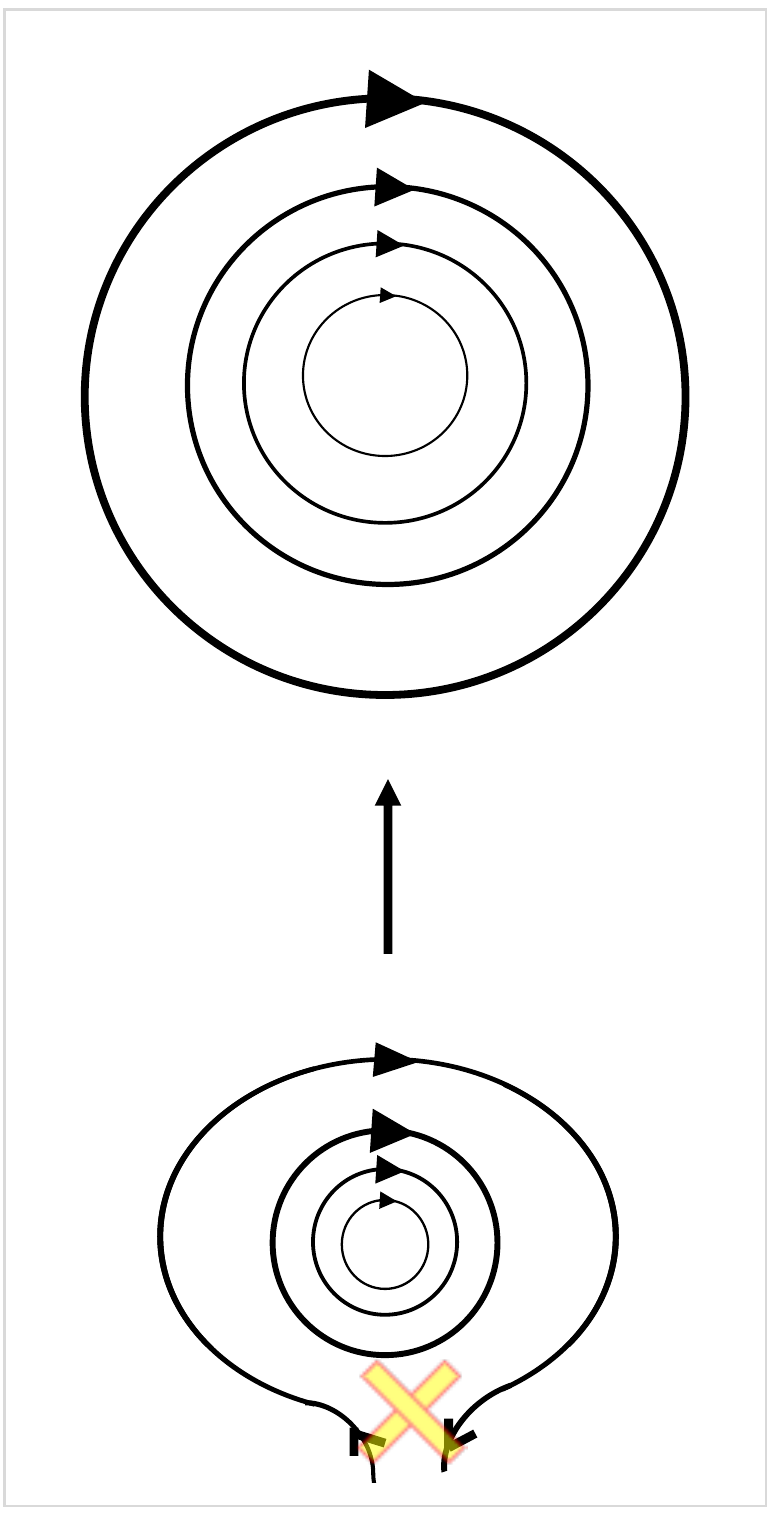}{0.3\textwidth}{(c)}
          }
\caption{A sketch of three different FR expansion trends: a) relative increase in the FR cross-sectional area is ``balanced" by the decrease in the strength of the central magnetic field (magnetic flux conserved); b) increase in the FR cross-sectional area is relatively slow compared to the decrease in the strength of the central magnetic field (magnetic flux decreased); c) increase in the FR cross-sectional area is relatively fast compared to the decrease in the strength of the central magnetic field (magnetic flux increased). View of the FR cross section is shown with field lines in poloidal direction. The yellow cross marks the point of reconnection (interaction with the ambient plasma).
\label{fig2}}
\end{figure*}

In Equation \ref{eq9} we assumed that the ratio $r(t)/a(t)$ does not change with time in order to apply the method of separation of variables. This assumption holds when the change rate of any shell within the cylinder is proportional to the change rate of its outermost shell, $\mathrm{d}r/\mathrm{d}a=const.$, \ie when $r(t)=const\cdot a(t)$. It can be easily shown that this holds when the cylinder is expanding self-similarly, \ie in a way that at later times it is a scaled reproduction of its original shape. This means that the coordinates of a plasma element at a given time are scaled by a time-dependent factor compared to the reference-time value $t_0$, $x_i(t)=x_i(t_0)f_i(t)$, where $x_i$ is the plasma coordinate and $f_i(t)$ a corresponding time-dependant factor in the $i$ direction \citep[][]{demoulin08}. Therefore, a self-similar expansion of a FR radius can be written as:

\begin{equation}
a(t) = a_0\Bigg(\frac{R(t)}{R_0}\Bigg)^{n_a}\,,
\label{eq13}
\end{equation}

\noindent where $R(t)$ is the heliospheric distance at time $t$, $R_0$ is the starting heliospheric distance, $a_0$ is the starting FR radius and $n_a$ is the power-law index which observational studies approximately constrain to $0.45<n_a<1.14$ \citep[see \eg][]{bothmer98,leitner07,demoulin08,gulisano12}. An expression similar to Equation \ref{eq13} can be used to describe the corresponding decrease of the central magnetic field:

\begin{equation}
B(t) = B_0\Bigg(\frac{R(t)}{R_0}\Bigg)^{-n_B}\,,
\label{eq14}
\end{equation}

\noindent where $B_0$ is the initial magnetic field and $n_B$ is the power-law index which observational studies approximately constrain to $0.88<n_B<1.89$ \citep[\eg][and references therein]{gulisano12}. Using Equations \ref{eq13} and \ref{eq14} we can consider different types of FR expansion based on the axial magnetic flux. The axial magnetic flux is given as $\Phi_{ax}=B_{\phi}S$, where $B_{\phi}$ is the axial magnetic field, which is in a force free FR model related to the magnetic field in the FR center, $B_c$ \citep[\eg][]{lundquist51},  and $S$ is the cross section of the FR. Assuming a circular cross section the axial magnetic flux is given by $\Phi_{ax}\sim B_ca^2$, which can be used to determine the power-law index in Equation \ref{eq14} with the assumption that the magnetic flux is conserved ($n_B=2n_a$). However, observational studies involving $n_B$ and $n_a$ measurements have shown that $n_B$ is not necessarily equal to $2n_a$ \citep[see][and references therein]{gulisano12}. For instance, \citet{gulisano10} report $n_a=0.89$ and $n_B=1.85$ for non-perturbed magnetic clouds (those showing pronounced linear velocity profile), \ie $n_B\approx2n_a$. However, for the perturbed magnetic clouds they found that they expand less with distance, with $n_a=0.45$ and $n_B=1.89$, \ie $n_B-2n_a>0$. On the other hand, \citet{leitner07} found that magnetic clouds in general show $n_a=1.14$ and $n_B=1.64$, \ie $n_B-2n_a<0$. Mentioned studies are focused on the inner solar system between 0.3 and 1 AU, which is the main focus of this study as well, therefore, based on these observational studies we consider three different expansion trends: $n_B-2n_a=0$, $n_B-2n_a>0$, and $n_B-2n_a<0$. Physically, these expansion trends relate $n_B-2n_a=0$ to the magnetic flux conservation, $n_B-2n_a>0$ to a decrease of the magnetic flux with heliospheric distance, whereas $n_B-2n_a<0$ is related to the increase of the magnetic flux. Observational evidence for the reduction of the magnetic flux through erosion due to magnetic reconnection was found in studies by \eg\, \citet{dasso07} and \citet{ruffenach15}, whereas \citet{manchester14} suggest that the reconnection at the rear of the flux rope can lead to flux injection \citep[for an overview of various processes affecting the axial magnetic flux evolution see][]{manchester17}. We note that the interaction of the FR with the ambient plasma affects the ordered structure of the FR in the outer FR layers, but from the aspect of the model we are only concerned about the relative increase of the FR cross-sectional area compared to the decrease of the central magnetic field strength. This is visualised by three corresponding sketches in Figure \ref{fig2}. 

Based on these considerations on the FR expansion we can determine the function $f(t)=\int D(t)/a(t)^2 \mathrm{d}t$, which defines the time dependent part of the solution given in Equation \ref{eq12}. We assume that the FR moves with a constant velocity $v=R/t$ and that the change of the FR radius and of the central magnetic field is given by Equations \ref{eq13} and \ref{eq14}, respectively. Furthermore, we assume that the diffusion coefficient relates to the magnetic field strength $D\sim1/B$ \citep[see \eg][and references therein]{potgieter13} and therefore increases with heliospheric distance with a power-law index $n_B$. With these assumptions the time dependent part is reduced to:

\begin{equation}
f(t)=\frac{D_0}{a_0^2}\cdot \Big(\frac{v}{R_0}\Big)^x\cdot \int t^x \mathrm{d}t\,,
\label{eq15}
\end{equation}

\noindent where $x=n_B-2n_a$. In the case of the conserved magnetic flux $x=0$ and integration is trivial. In cases where the flux is not conserved we consider simple options where $x=0.5$ and $x=-0.5$ for decreased and increased flux, respectively. The selected options are simple to integrate and are related to the $n_B$ and $n_a$ ranges restricted by the above mentioned observational studies. For both $x=0.5$ and $x=-0.5$ integration results in rational functions, as well as for any other value of $x$, except for $x=-1$ when integration of $f(t)\sim \int t^x \mathrm{d}t$ results in a logarithmic function. Therefore, $x=-1$ is considered as a fourth, special case of expansion. We note that reconnection would change the magnetic connectivity of the FR allowing additional GCRs to stream in \citep{masias-meza16}. Therefore, it is reasonable to expect that the diffusion coefficient would be influenced to some extent. On the other hand, since $f(t)$ and $x$-values we use in the model are arbitrarily selected  (although empirically based), for the sake of simplicity, we do not include this effect in our calculations.

To summarise, we consider the solution of the diffusion-expansion equation given in Equation \ref{eq12} for four different types of expansion which lead to four different types of $f(t)$ governing the time-behaviour of the modelled FD:

\begin{enumerate}
\item $n_B-2n_a=0, x=0 \longrightarrow f(t)\propto t$ (conserved magnetic flux)
\item $n_B-2n_a>0, x=0.5 \longrightarrow f(t)\propto t^{\frac{3}{2}}$ (reduced magnetic flux)
\item $n_B-2n_a<0, x=-0.5 \longrightarrow f(t)\propto t^{\frac{1}{2}}$ (increased magnetic flux)
\item $n_B-2n_a<0, x=-1 \longrightarrow f(t)\propto \ln{(at+1)}$ (increased magnetic flux - special case)
\end{enumerate}

We note that initial CME parameters, which are needed to quantify the solution of Equation \ref{eq15} ($a_0$, $v$, and $R_0$), can be obtained from remote CME observation using \eg\, 3D CME reconstruction. On the other hand, $D_0$ is unknown and cannot be obtained from remote observation, therefore it has to be estimated. For that purpose we estimate the diffusion coefficient near Earth, $D_E$ \citep[following the typical empirical expression used in numerical models, see][]{potgieter13} and back-extrapolate to $R_0=20$\Rsun using assumed power-law behaviour. Namely, we assume $D\sim 1/B$, $B$ following the power-law given in Equation \ref{eq14}, where $n_B=2$ for expansion types $x=0$ and $x=0.5$, and $n_B=1$ for expansion types $x=-0.5$ and $x=-1$. Given that $x=n_B-2n_a$, this implicitly defines the power-law index given in Equation \ref{eq13}, where $n_a=1$ for expansion type $x=0$, $n_a=0.75$ for $x=0.5$, $n_a=0.75$ for $x=-0.5$, and $n_a=1$ for $x=-1$.

\begin{sidewaysfigure}
\centerline{\includegraphics[width=0.24\textwidth]{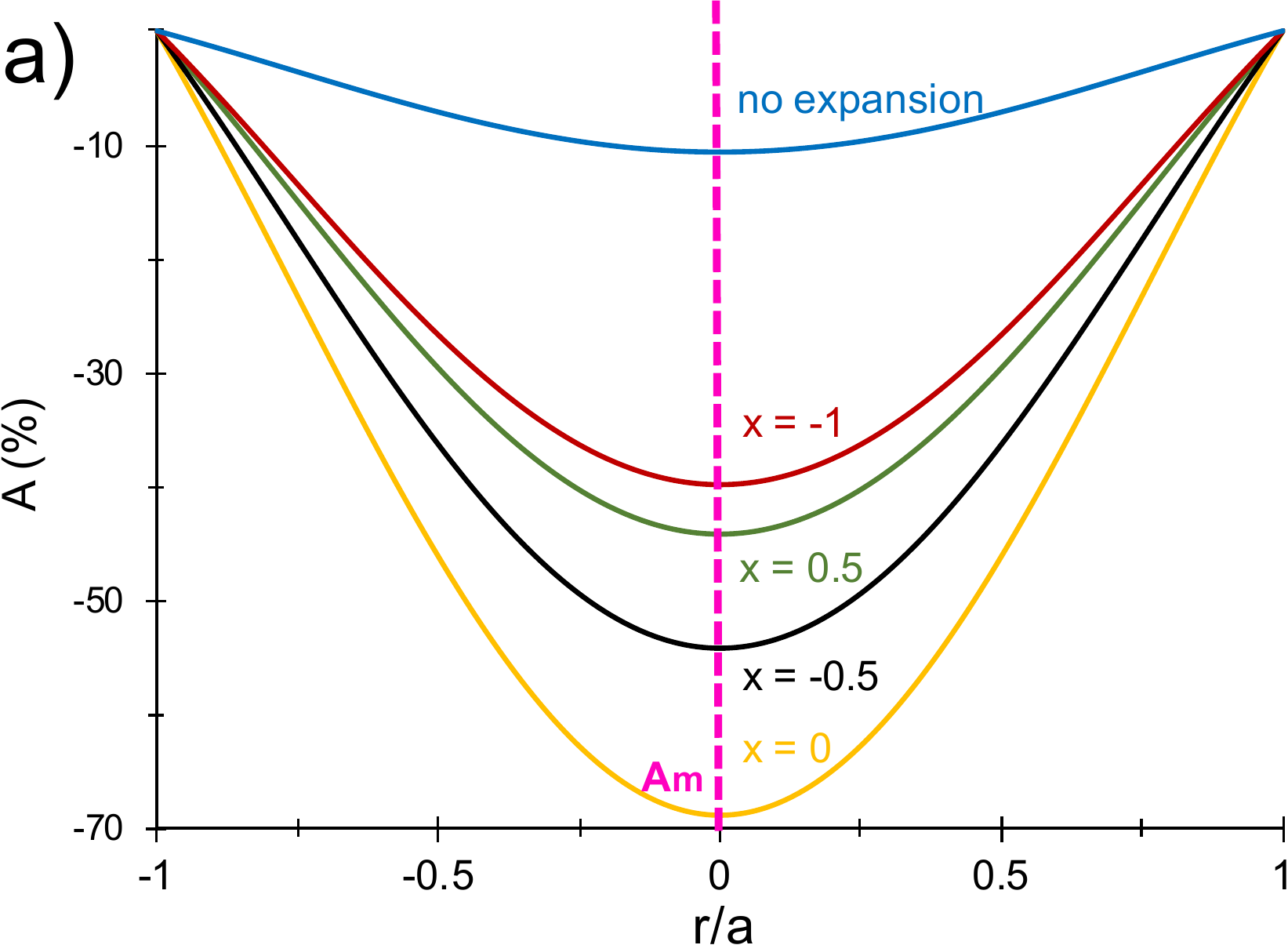}
		\includegraphics[width=0.24\textwidth]{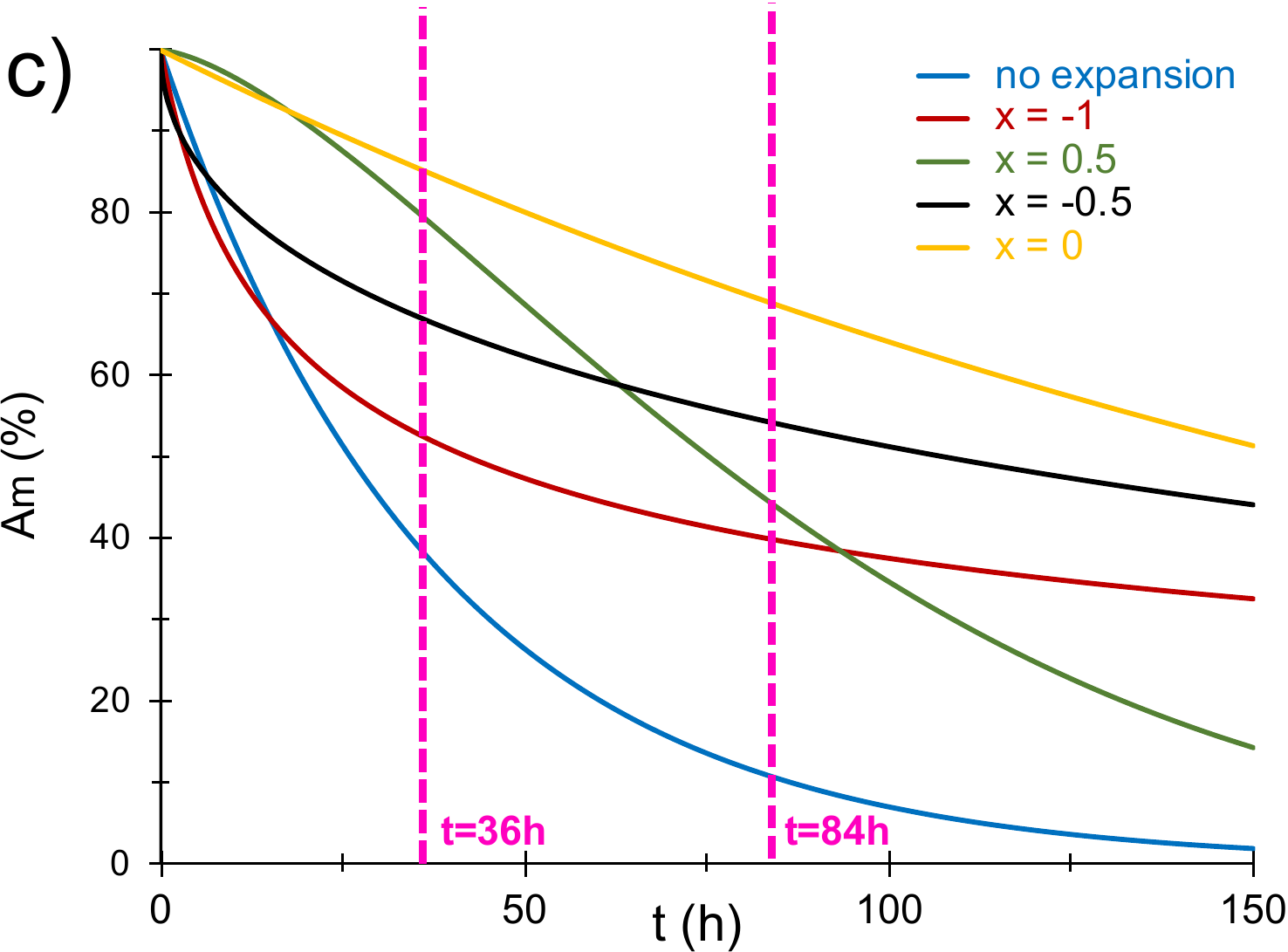}
		\includegraphics[width=0.24\textwidth]{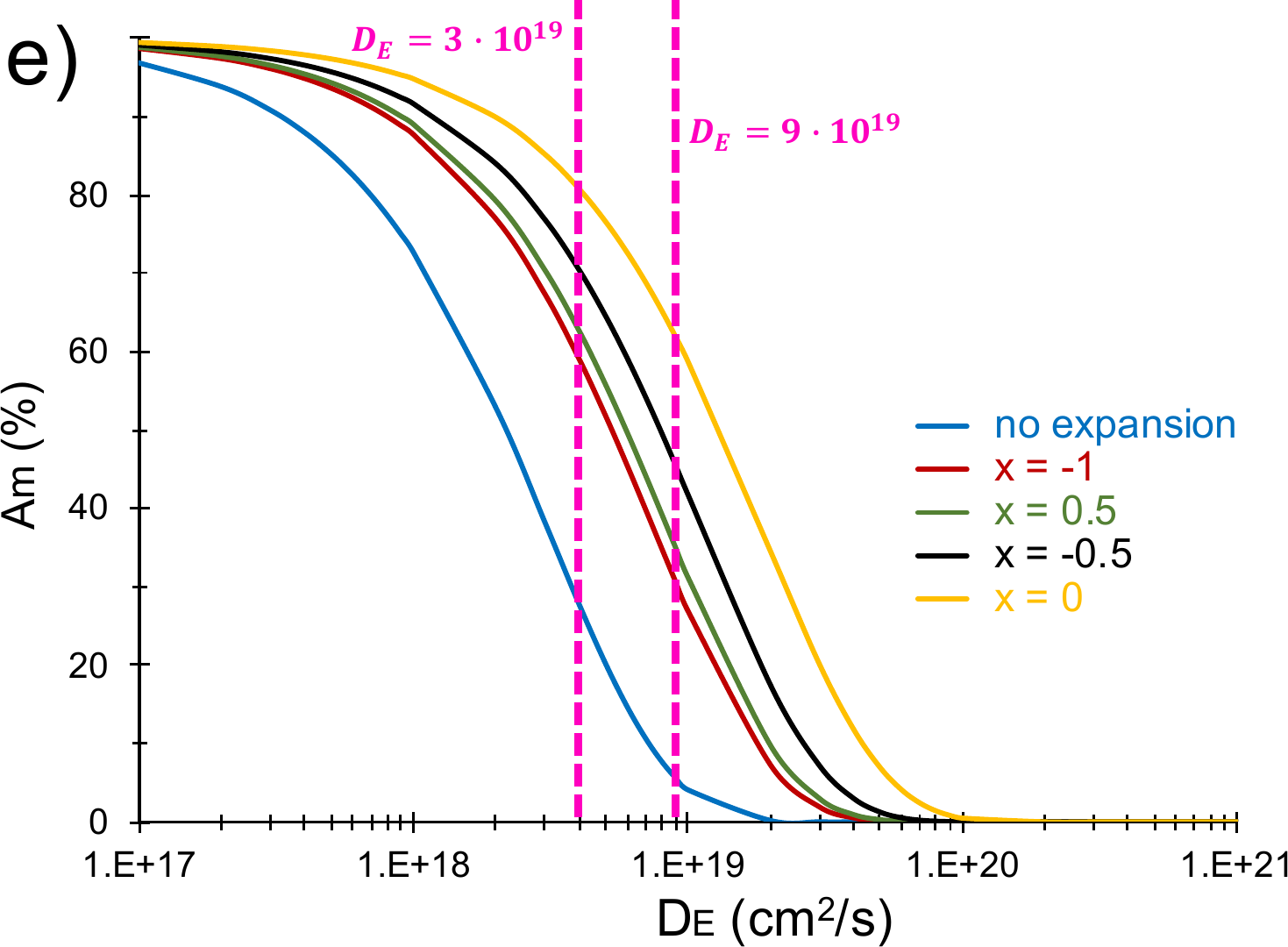}
		\includegraphics[width=0.24\textwidth]{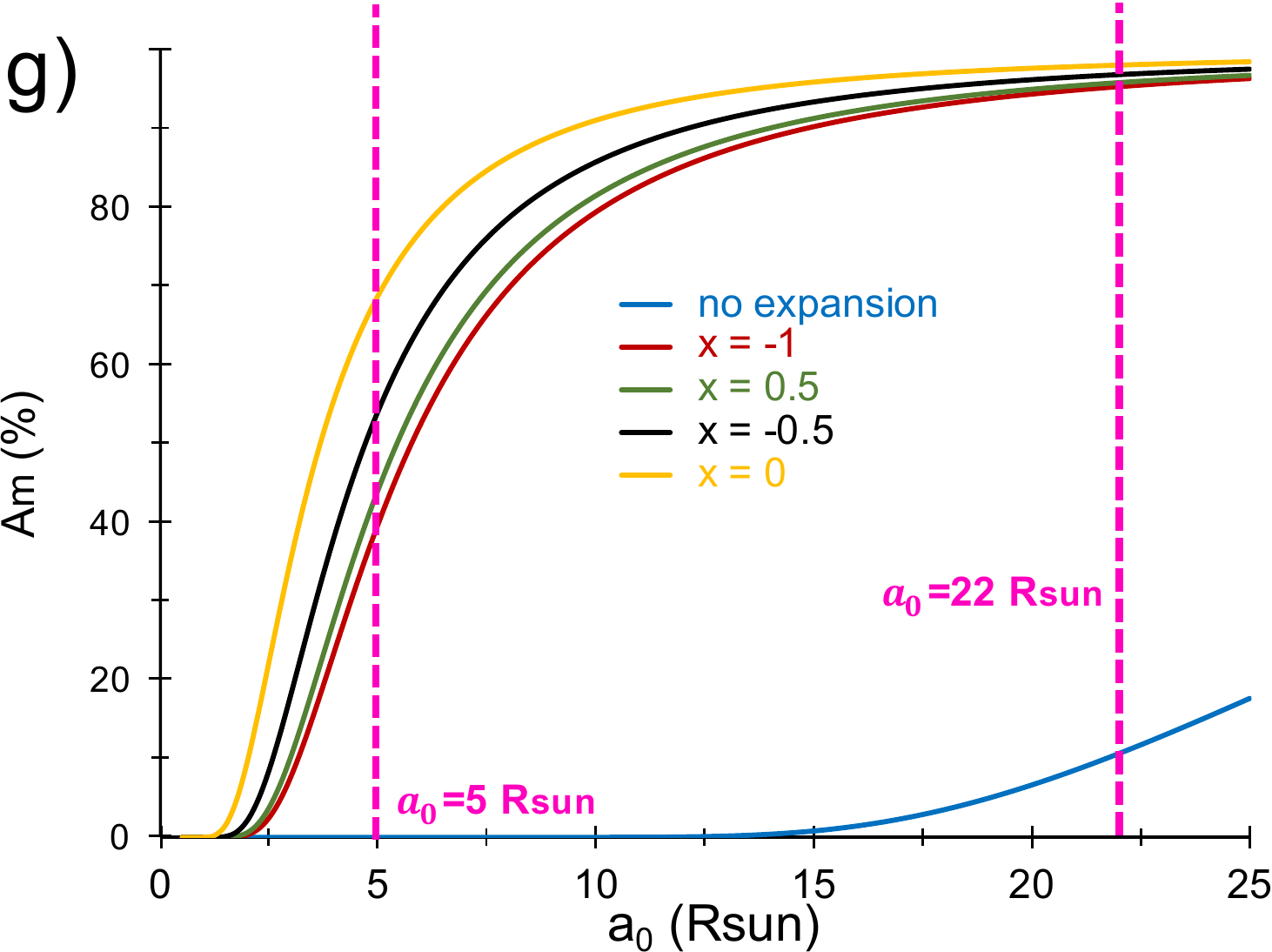}}
\vspace{0.03\textwidth}
\centerline{\includegraphics[width=0.24\textwidth]{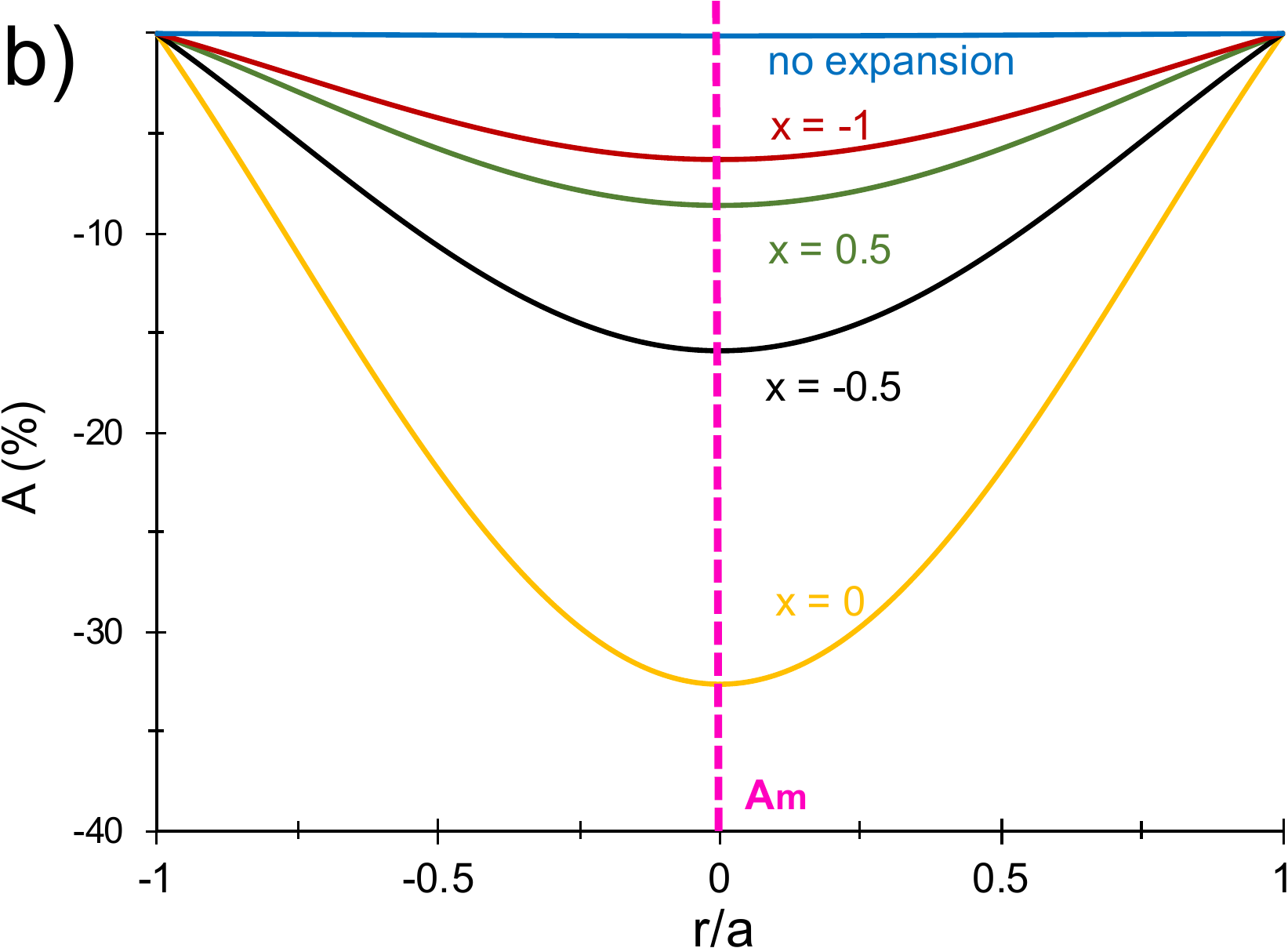}
		\includegraphics[width=0.24\textwidth]{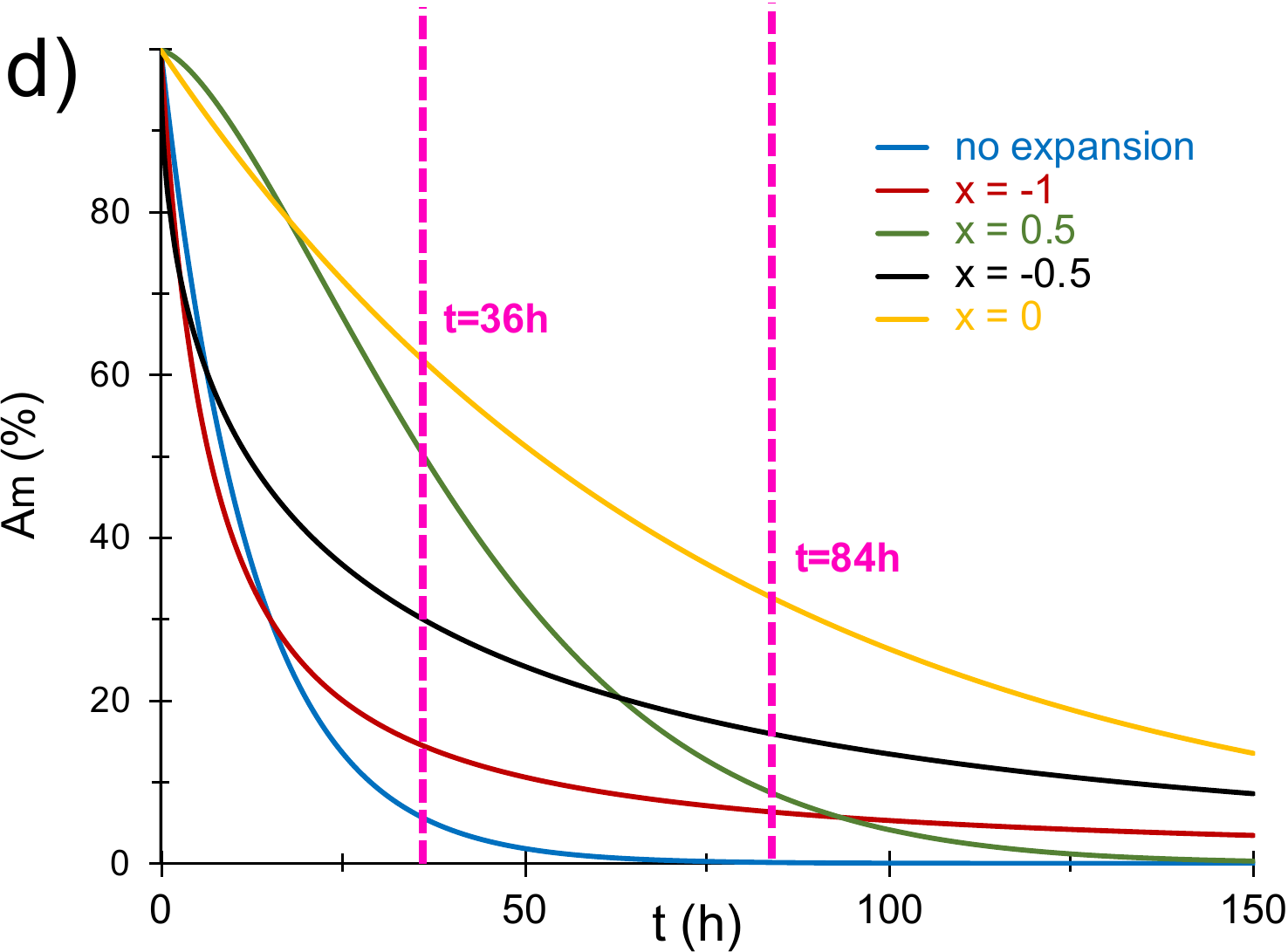}
		\includegraphics[width=0.24\textwidth]{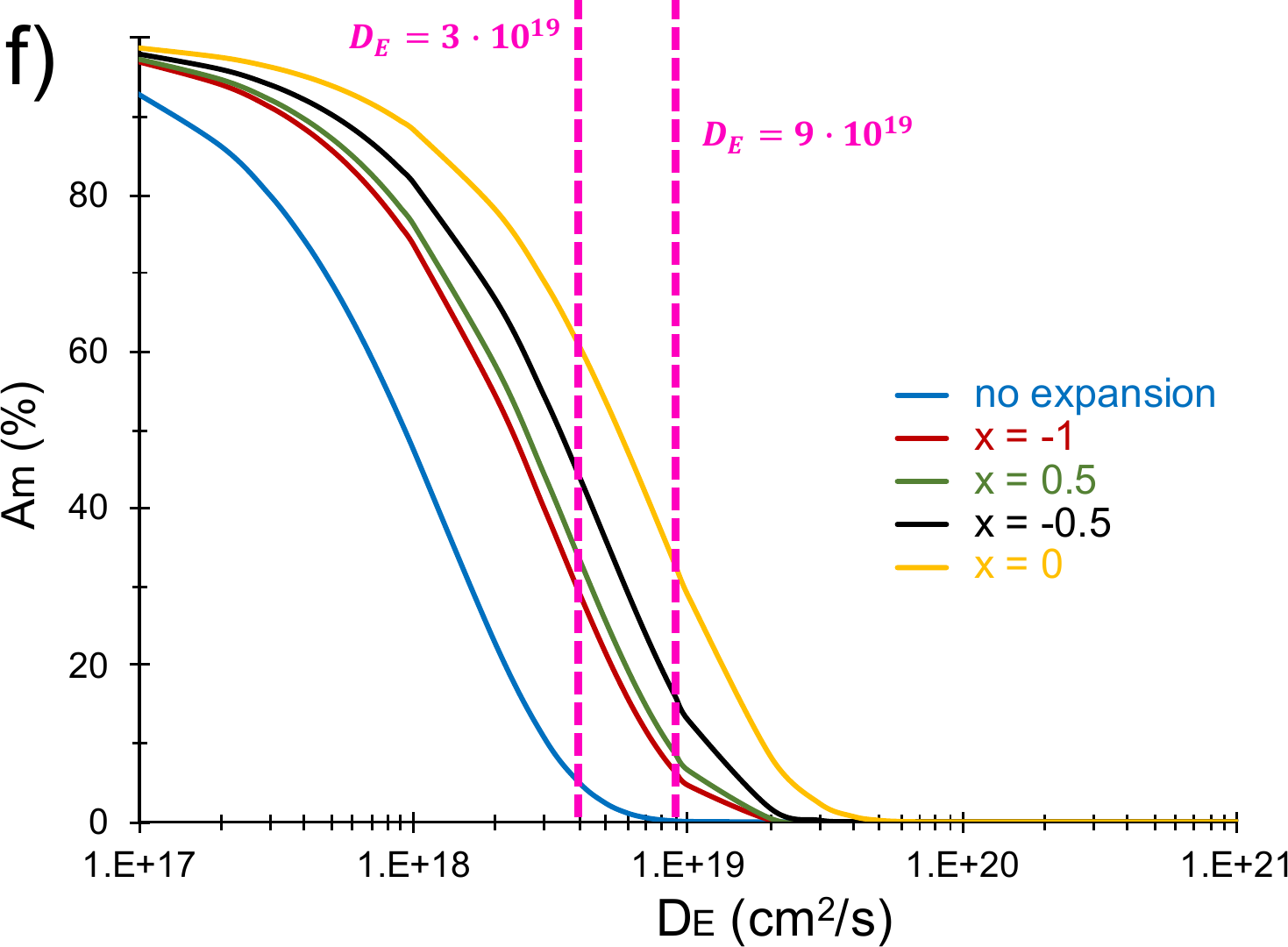}
		\includegraphics[width=0.24\textwidth]{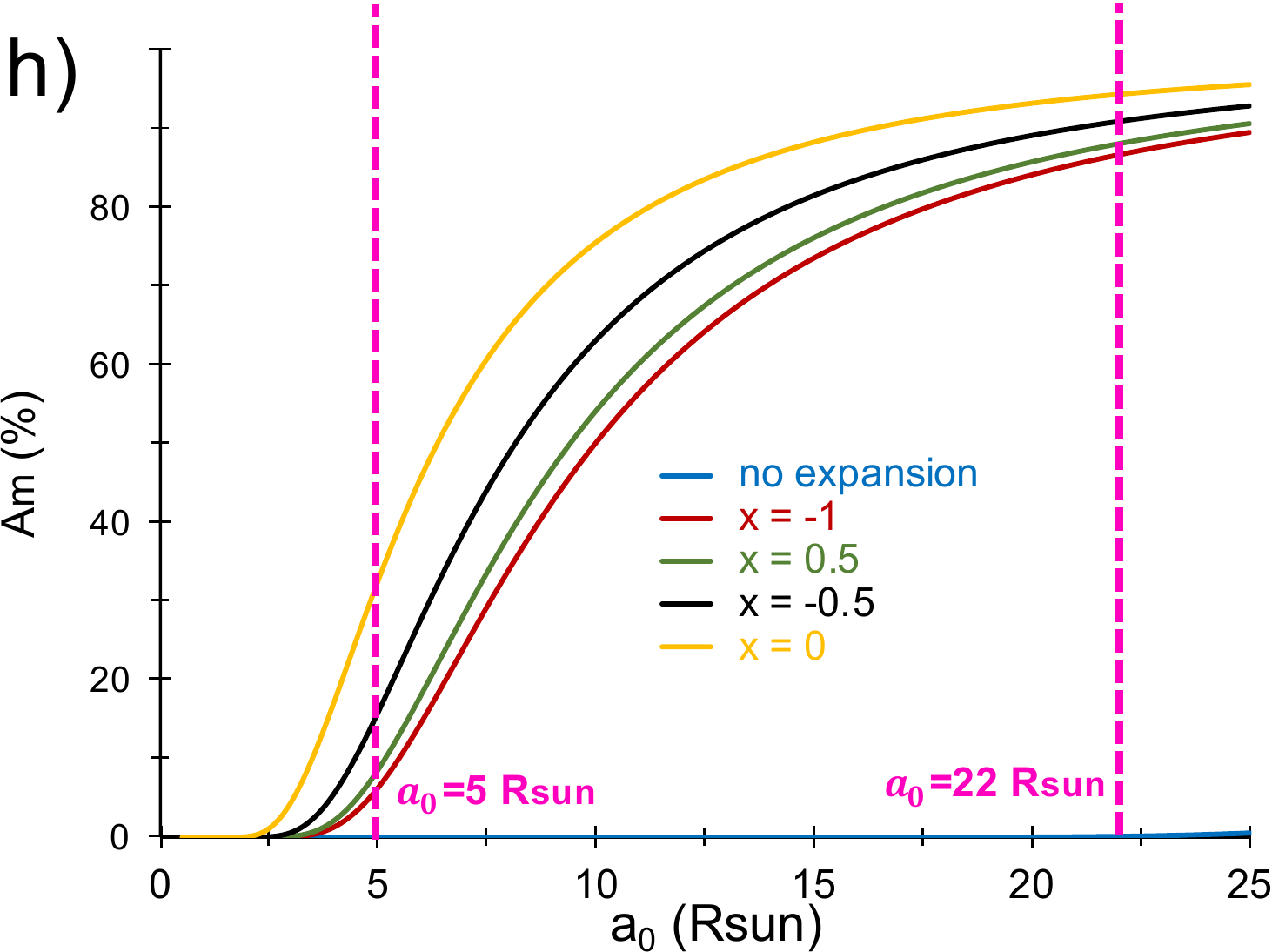}}
\vspace{0.03\textwidth}
\caption{
a) The relative GCR phase space density, $A$, as a function of the radial position within the FR, $r/a$, after a transit time of $84$ h for different types of expansion of a generic CME ($a_0=5$\Rsun, $R_0=20$\Rsun, $v\approx500$\kmps) and no expansion case ($a=0.1$ AU, $D_0=D_E$) for $D_E=3\times10^{18}$\Dunit ($D_0=2.6\times10^{16}$\Dunit\, for $x=0$ and $x=0.5$; $D_0=2.8\times10^{17}$\Dunit\, for $x=-0.5$ and $x=-1$). The dashed magenta line marks the position of the FR center where the relative FD magnitude, $A_{\mathrm{m}}$, is defined.
\newline b) Same as in a) for $D_E=9\times10^{18}$\Dunit ($D_0=7.7\times10^{16}$\Dunit\, for $x=0$ and $x=0.5$; $D_0=8.4\times10^{17}$\Dunit\, for $x=-0.5$ and $x=-1$).
\newline c) Time evolution of the FD magnitude, $A_{\mathrm{m}}$ for $D_E=3\times10^{18}$\Dunit. The same generic CME and different expansion types are applied as in a) and b). Two selected transit times are highlighted by dashed magenta lines.
\newline d) Same as in c) for $D_E=9\times10^{18}$\Dunit.
\newline e) FD magnitude at Earth, $A_{\mathrm{m}}$, \textit{vs} diffusion coefficient at Earth, $D_E$, for a generic CME ($a_0=5$\Rsun\,,$R_0=20$\Rsun\,) moving at a speed of $\approx1150$ \kmps\,(transit time 36 h). Different expansion types are applied as in a)--d). Two selected $D_E$ are highlighted by dashed magenta lines.
\newline f) Same as in e) for $v=500$\kmps\, (transit time 84 h).
\newline g) FD magnitude, $A_{\mathrm{m}}$, after a transit time of $84$ h \textit{vs} the initial FR radius, $a_0$, for $D_E=3\times10^{18}$\Dunit\, and different expansion types as in a) and b). Two selected $a_0$ are highlighted by dashed magenta lines.
\newline h) Same as in g) for $D_E=9\times10^{18}$\Dunit\,.}
\label{fig3}
\end{sidewaysfigure}

In Figure \ref{fig3} the solutions of the model for different types of expansion, diffusion coefficients, transit times and FR radius are presented. Figures \ref{fig3}a and b show relative GCR phase space density as a function of the radial position within the FR after a transit time of $84$ h for two different values of the initial diffusion coefficient, $D_0$, respectively. Different $D_0$ were estimated as described above based on the diffusion coefficients at Earth, $D_E$, which roughly correspond to values obtained from Equations 23 and 24 by \citet{potgieter13} for 1 GV GCRs in a 10 and 30 nT magnetic field. A generic CME was used, having the initial radius $a_0=5$\Rsun\, at a distance $R_0=20$\Rsun, and arriving at Earth after a transit time of $t=84$ h (\ie moving at a constant speed of $\approx500$\kmps). The dashed magenta line marks the center of FR, defining the relative FD magnitude, $A_{\mathrm{m}}$. The five curves correspond to different types of expansion: no expansion (diffusion only approach with constant size, as described in Section \ref{model1}) and four different types of expansion described above. In the case of no expansion the radius of the generic CME is constant and has a value $a=0.1$ AU, and the diffusion coefficient is also constant with a value corresponding to $D_E$. Figures \ref{fig3}c and d show the time evolution of the FD magnitude, $A_{\mathrm{m}}$, for two different values of $D_0$ used in Figures \ref{fig3}a and b, respectively. Also the same generic CME and different expansion types are applied as in Figures \ref{fig3}a and b. Two selected transit times are highlighted by dashed magenta lines. Figures \ref{fig3}e and f show the FD magnitude at Earth, $A_{\mathrm{m}}$, \textit{vs} the diffusion coefficient at Earth, $D_E$, for a generic CME ($a_0=5$\Rsun, $R_0=20$\Rsun) moving at a speed of $\approx1150$ and $\approx500$\kmps, respectively (corresponding to the transit times to Earth of 36 and 84 h, respectively). Different expansion types are applied as in Figures \ref{fig3}a-d. Two selected $D_E$ (used in Figures \ref{fig3}a and b) are highlighted by dashed magenta lines. Finally, Figures \ref{fig3}g and h show the FD magnitude, $A_{\mathrm{m}}$, after a transit time of $84$ h \textit{vs} the initial FR radius for two different values of $D_0$ used in Figures \ref{fig3}a and b, respectively. Two selected initial FR radii ($a_0=0.5$\Rsun and 0.1 AU used in Figures \ref{fig3}a and b for different expansion types and no expansion, respectively) are highlighted by dashed magenta lines (for no expansion case initial FR radius equals the one at Earth because $a=const.$).

It can be seen in Figures \ref{fig3}a and b that the relative GCR phase space density has a symmetric shape within the FR, reaching its minimum in the center of the FR and is restricted to the FR extent. The shape is qualitatively the same regardless of the expansion type and diffusion coefficient and shows a qualitative agreement with observed ejecta-FD profiles \citep[\eg][]{cane93,belov15,masias-meza16}. Furthermore, we note that qualitatively, for all expansion cases, $A_{\mathrm{m}}$ decreases with time as can be seen in Figures \ref{fig3}c and d, which is also in agreement with observational studies \citep{cane94,blanco13b}. Quantitatively, it can be seen in Figures \ref{fig3}a and b that the amplitude of the depression depends on both diffusion coefficient and expansion type. The relative amplitude of the GCR phase space density in the FR center, \ie the FD magnitude, $A_{\mathrm{m}}$, is smaller for larger diffusion coefficient at Earth, $D_E$, which is also evident from Figures \ref{fig3}e and f. This reflects the fact that larger $D_E$ means more efficient diffusion of GCRs into the FR, \ie faster filling up. Therefore, for the same diffusion time and $a_0$, FR with larger $D_E$ will have more GCRs, \ie smaller $A_{\mathrm{m}}$ than the FR with smaller $D_E$.

\begin{figure*}
\gridline{\fig{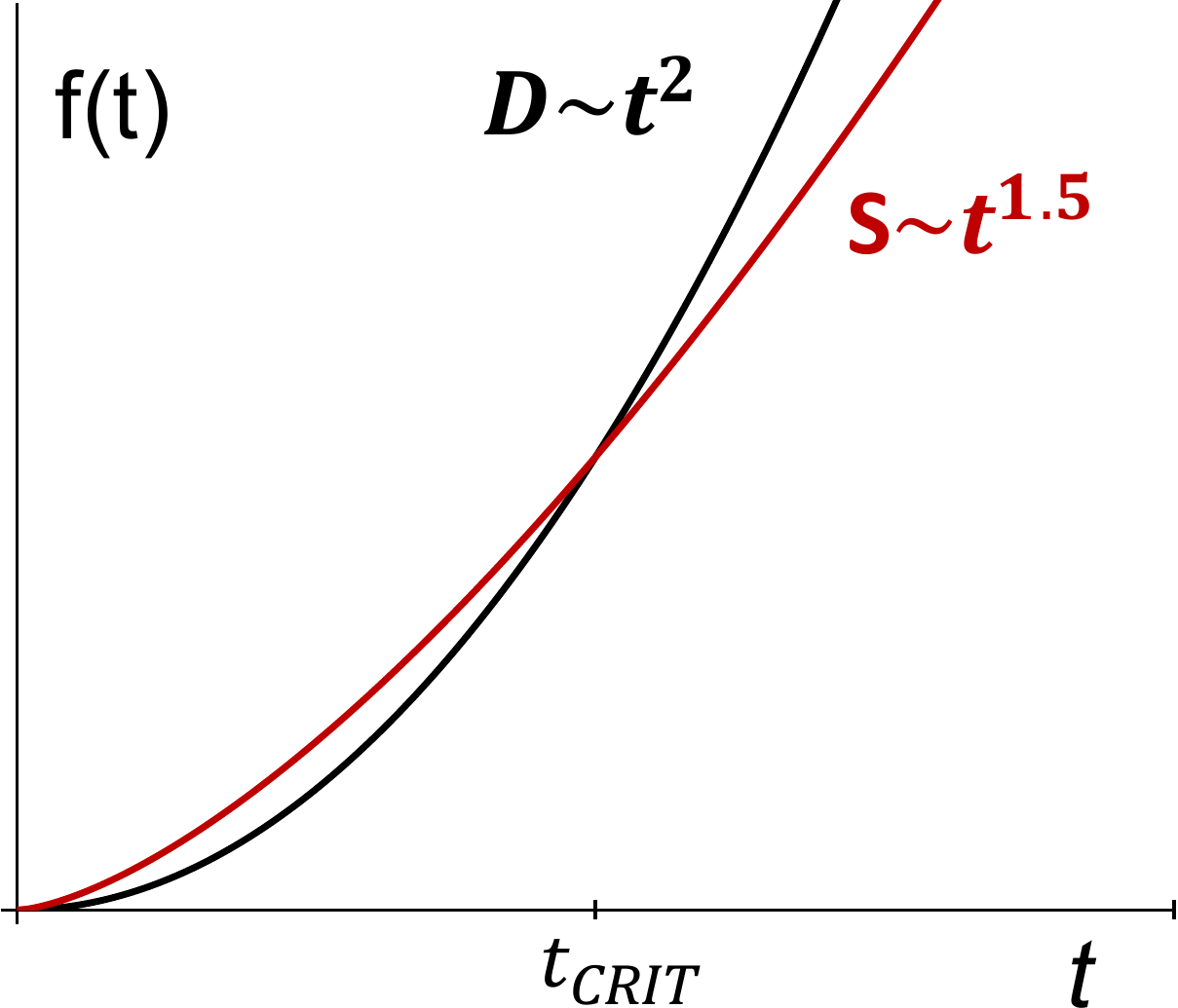}{0.22\textwidth}{(a)}
          \fig{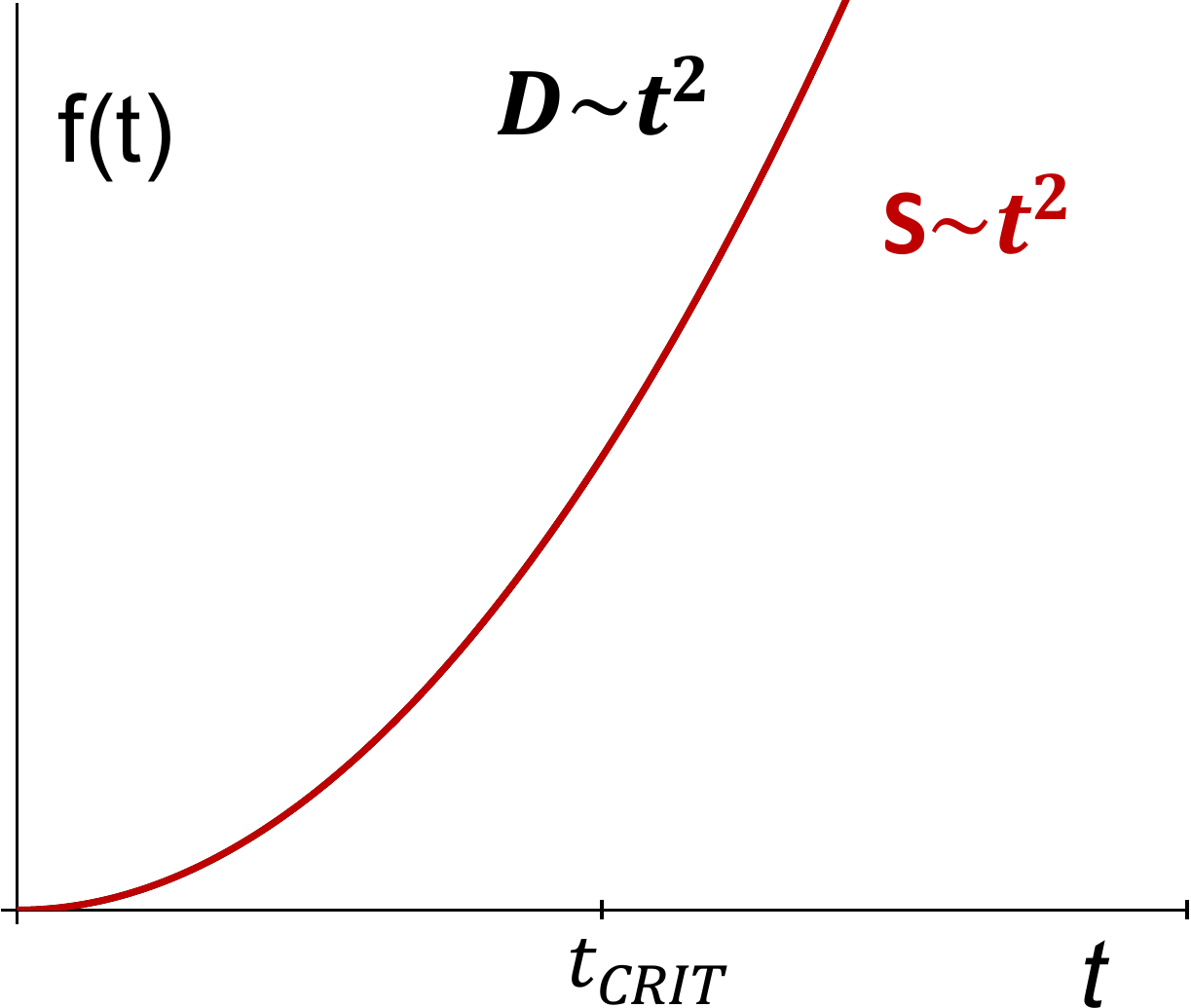}{0.22\textwidth}{(b)}
          \fig{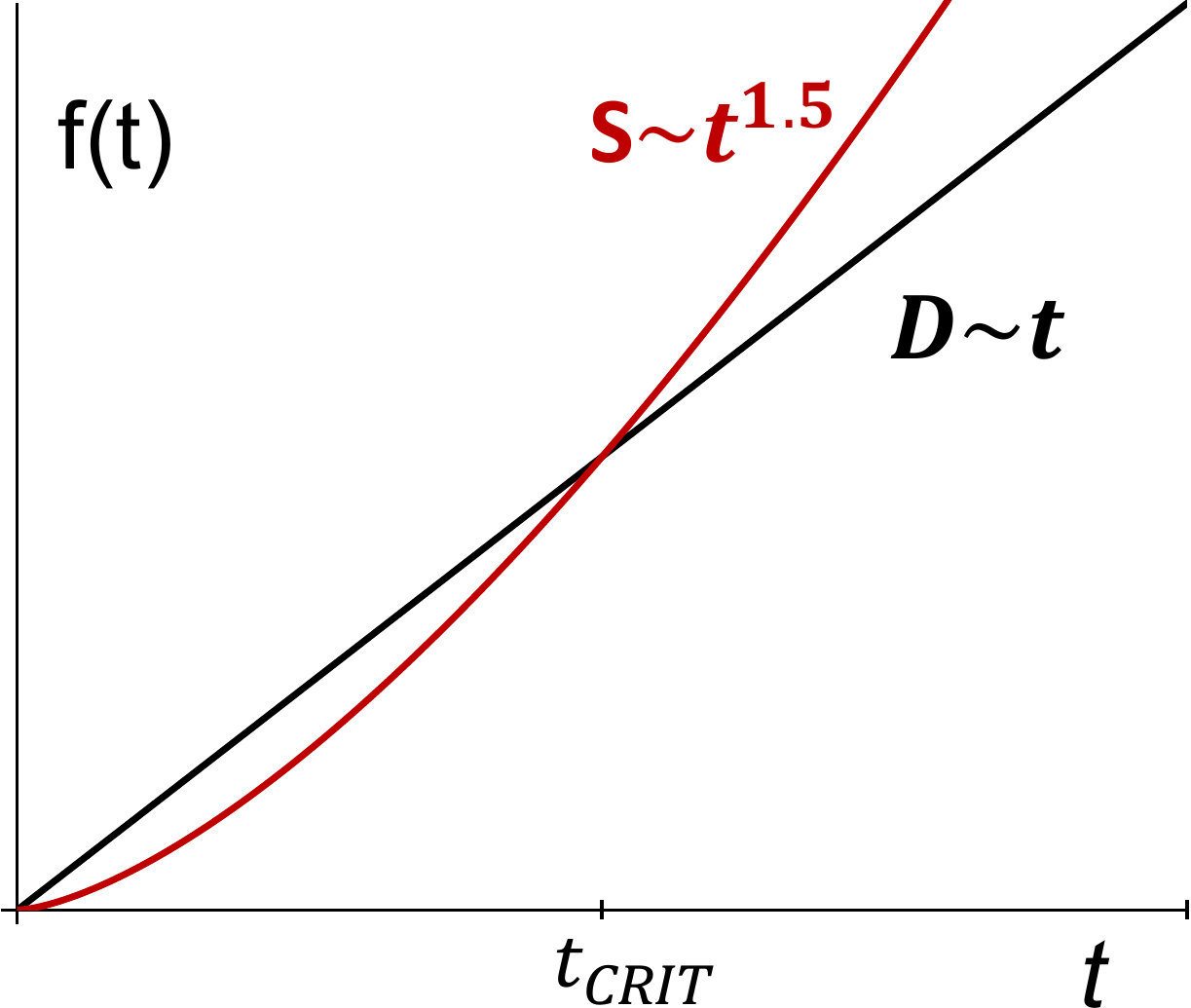}{0.22\textwidth}{(c)}
          \fig{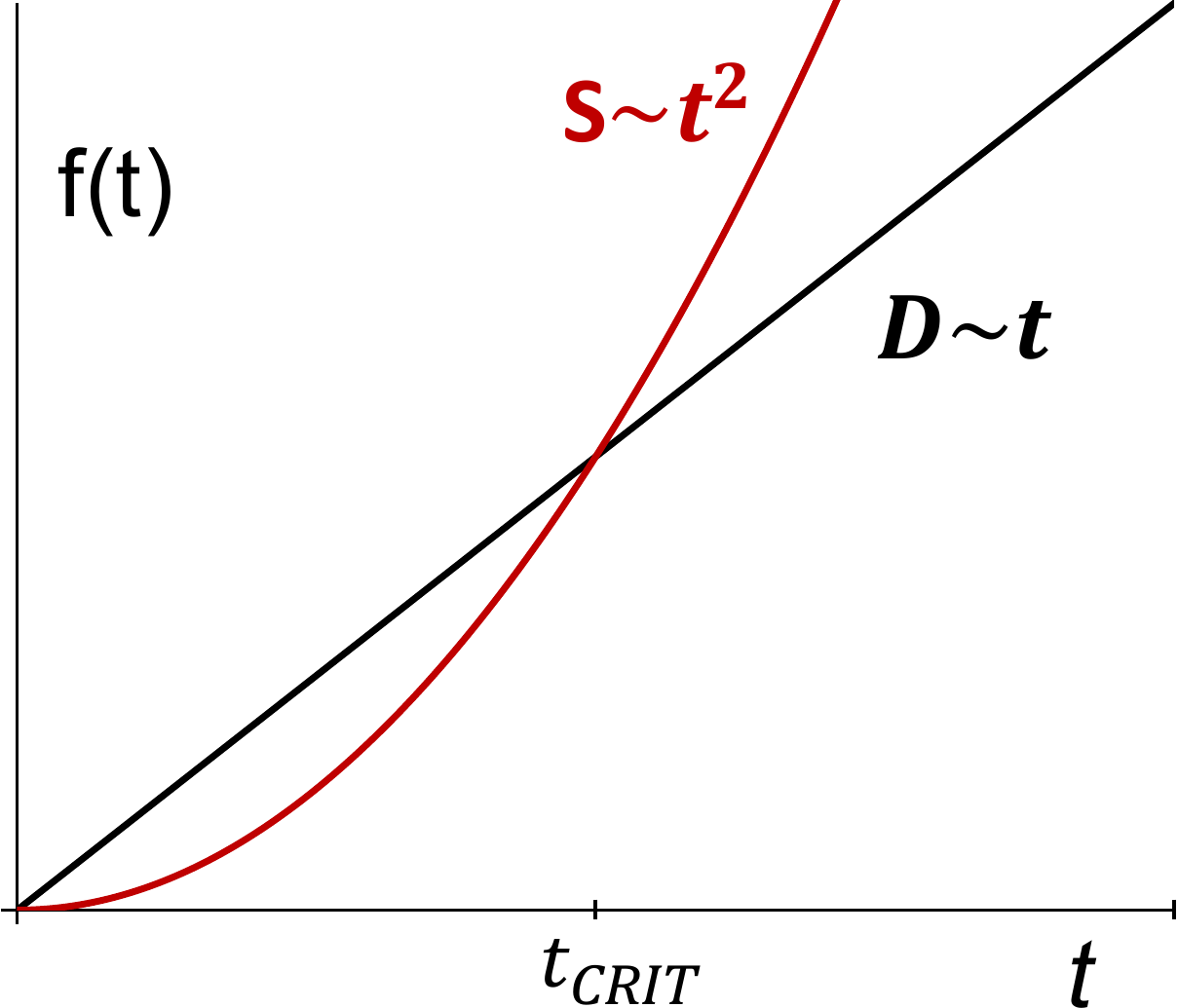}{0.22\textwidth}{(d)}
          }
\caption{Time dependencies of diffusion coefficient, $D$, and cross-sectional area, $S$, for different expansion types: a) $x=0.5$ (magnetic flux decreased); b) $x=0$ (magnetic flux conserved); c) $x=-0.5$ (magnetic flux increased); d) $x=-1$ (magnetic flux increased - special case). $D$, $S$, and $t$ are given in arbitrary units. $t_{CRIT}$ marks the point between small and large $t$--regimes.
\label{fig4}}
\end{figure*}

The relation to the expansion type is not so simple. Comparing Figures \ref{fig3}a and b one might get the impression that $A_{\mathrm{m}}$ is smallest for the no expansion case and largest for the $x=0$ expansion type. However, Figures \ref{fig3}c and d reveal that this depends strongly on the transit time after which $A_{\mathrm{m}}$  is observed. In general, we expect that $A_{\mathrm{m}}$ depends on a complex interplay of diffusion and expansion as given by Equation \ref{eq15}. However, it should be noted that these two effects are not independent of each other. On one hand, the expansion increases the diffusion coefficient (as $B$ decreases) enhancing the diffusion, while on the other hand, increases the FR size and therefore acts as a ``diluting" mechanism (see Figures \ref{fig3}g and h). We also note that the initial diffusion coefficients $D_0$ are not the same for all expansion types, due to the fact that we fix the diffusion coefficient at Earth $D_E$ and back-extrapolate it using different power-laws. Finally, the functional character of the solutions obtained from Equation \ref{eq15} needs to be taken into account. This is shown in Figure \ref{fig4}, where for different expansion types the time-dependencies of diffusion coefficient, $D$, and cross-sectional area, $S$, are shown (in arbitrary units).

It can be seen in Figure \ref{fig4} that the expansion can both enhance and slow down the diffusion for different expansion types, depending on the time regime, which is defined according to $t_{CRIT}$ (determined by the values of $D$, $S$, and $t$).

Namely, in the small t-regime, $t<t_{CRIT}$, the following is observed for different expansion types:
\begin{itemize}
\item for $x>0$ the increase of $S$ is faster than the increase of $D$ $\longrightarrow$ the expansion slows down the diffusion
\item for $x=0$ the increase of $S$ is balanced with the increase of $D$ $\longrightarrow$ the expansion does not affect the diffusion
\item for $x<0$ the increase of $S$ is slower than the increase of $D$ $\longrightarrow$ the expansion enhances the diffusion
\end{itemize}

Conversely, in the large t-regime, $t>t_{CRIT}$, we observe the following for different expansion types:
\begin{itemize}
\item for $x>0$ the increase of $S$ is slower than the increase of $D$ $\longrightarrow$ the expansion enhances the diffusion
\item for $x=0$ the increase of $S$ is again balanced with the increase of $D$ $\longrightarrow$ the expansion does not affect the diffusion
\item for $x<0$ the increase of $S$ is faster than the increase of $D$ $\longrightarrow$ the expansion slows down the diffusion
\end{itemize}

Therefore, in different transit time regimes, different expansion types show different filling efficiency. This results in complex trends of the $A_{\mathrm{m}}(t)$ curves for different expansion types, which cross each other on several occasions (see Figure \ref{fig3}c and d). For example, at very small transit times ($t<3$h in Figure \ref{fig3}c) the largest $A_{\mathrm{m}}$ is for $x=0.5$ (magnetic flux reduced), whereas the smallest $A_{\mathrm{m}}$ is for $x=-1$ (magnetic flux increased-special case). At larger transit times ($t>18$h in Figure \ref{fig3}c) the largest $A_{\mathrm{m}}$ is for $x=0$ (magnetic flux conserved), whereas the smallest $A_{\mathrm{m}}$ is for no expansion case. We note that after some critical transit time (determined by initial conditions) the case of no expansion has smaller $A_{\mathrm{m}}$ than any expansion type. This is not surprising as for no expansion case a constant size assumption is used. For all expansion cases and for small $t$, FRs are relatively small compared to the no expansion case and thus easier to fill according to Figures \ref{fig3}g and h. At later stages, where the size of expanding FRs is comparable to the constant size FR, as there is no mechanism to counteract diffusion for no expansion case, the FR fills up more quickly resulting in a smaller $A_{\mathrm{m}}$.

In conclusion, at a given heliocentric distance, \ie after a specific transit time, $A_{\mathrm{m}}$ will be determined by a complex interplay of the diffusion and expansion depending on the initial conditions ($a_0$, $D_0$), as well as on the expansion type (\ie competing between the drop of $B$ and the increase of $S$).

Finally, we estimate how much the assumption that the GCR density outside the FR is constant, $U_0=const.$, influences our result. For that purpose we use the generic CME from Figure \ref{fig3}d and we assume that $U_0$ changes by $3\%$ throughout the evolution of the FR to 1 AU (within 84 h transit time), \ie $0.04\%$ per hour. We divide the FD amplitude evolution into 10 quasi-stationary time steps (each time-step lasts for 8.4 h), where between each time-step $U_0$ increases by $\approx0.3\%$. We assume that the increase of $U_0$ leads to a larger difference between FD amplitudes in two consecutive time-steps by factor $\alpha$: $\Delta A_{\mathrm{new,n}}=\alpha\cdot\Delta A_{\mathrm{old,n}}$, where $\Delta A_{\mathrm{old,n}}=A_{\mathrm{old,n}}-A_{\mathrm{old,n+1}}$ and $\alpha=1+0.3/100=1.003$ (``new" and ``old" correspond to $U_0\neq const.$ and $U_0=const.$, respectively). Here, the values of $A_{\mathrm{old}}$ was calculated based on Figure \ref{fig3}d, whereas $A_{\mathrm{new}}$ is calculated as $A_{\mathrm{new,n+1}}=A_{\mathrm{new,n}}-\Delta A_{\mathrm{new,n}}$ (in the first time step $A_{\mathrm{old,0}}=A_{\mathrm{new,0}}=1$). This procedure reflects the fact that in each time step there are more particles available to enter the FR compared to $U_0=const.$ case, but only a fraction of them actually enters the FR in the given time step. With this procedure we estimate a decrease in FD amplitude of 0.2--0.3\% (depending on the expansion type) for a 3\% increase of $U_0$ in a transit time of 84 h. This FD decrease corresponds to $< 5$\% relative uncertainty for the generic CME in Figure \ref{fig3}, which is reasonably low to be neglected. For higher values of the radial gradient, \eg, increase of $U_0$ of 10\%/AU, the estimated change in the FD amplitude is $\approx1\%$, corresponding to $\sim15\%$ relative uncertainty, too high to be neglected. However, we note that for the $\sim1$ GV protons, which are the main contributors to the observed FD in spacecraft, we expect the radial gradient to be $\sim3$\% \citep{gieseler16}, thus we expect the $U_0=const.$ assumption to hold.

Furthermore, we can apply the same procedure to estimate whether or not the model could be used to simulate the ejecta part during two-step FDs by assuming that the presence of shock/sheath can be introduced as the variation of $U_0$, where $U_0$ due to the reduced CR count rate in the sheath is $\approx5\%$ lower after 84 h \citep[based on the typical shock/sheath associated FD amplitude determined by][]{richardson11a}. We assume that the decrease in $U_0$ leads to a smaller difference between FD amplitudes in two consecutive time-steps by factor $\alpha=1-0.05/100=0.995$. With this procedure we estimate that a $5\%$ decrease of $U_0$ in a transit time of 84 h would lead to a decrease in FD amplitude of $<0.5\%$ for all expansion types, corresponding to $<8$\% relative uncertainty for the generic CME in Figure \ref{fig3}, which is still reasonably low. It should be noted, when the sheath region, \ie the standoff distance between CME and shock, increases with the radial distance, the variation of $U_0$ could depend much stronger on the radial distance than we assumed. In this case the variation rate of $U_0$ is much smaller than $\approx5\%/\mathrm{AU}$ at smaller radial distances and thus our estimation based on a constant variation rate can be regarded as an upper limit. We also note that in most ICMEs observed at Earth the shock/sheath part of FD amplitudes is $A_{\mathrm{shock}}\leq5\%$ \citep{richardson11a}. Therefore, based on our estimation, we conclude that for a large subset of CMEs the model could also be applied to explain the ejecta part in the case of the two-step FD.

\section{Case study: 2014 May 25 CME and 2014 May 30 Forbush decrease}
\label{case study}

\begin{figure}
\gridline{\fig{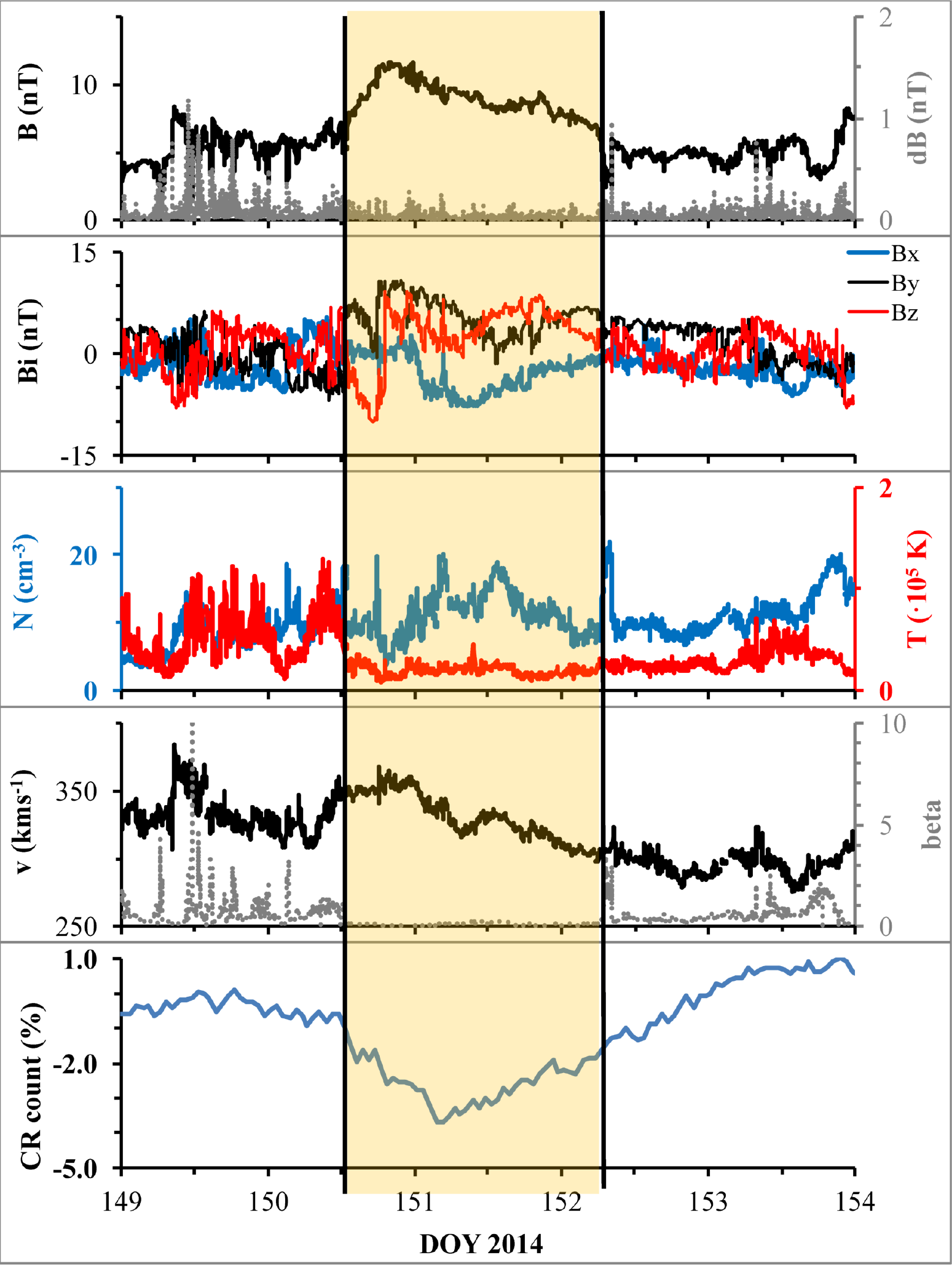}{0.45\textwidth}{(a)}
	      \fig{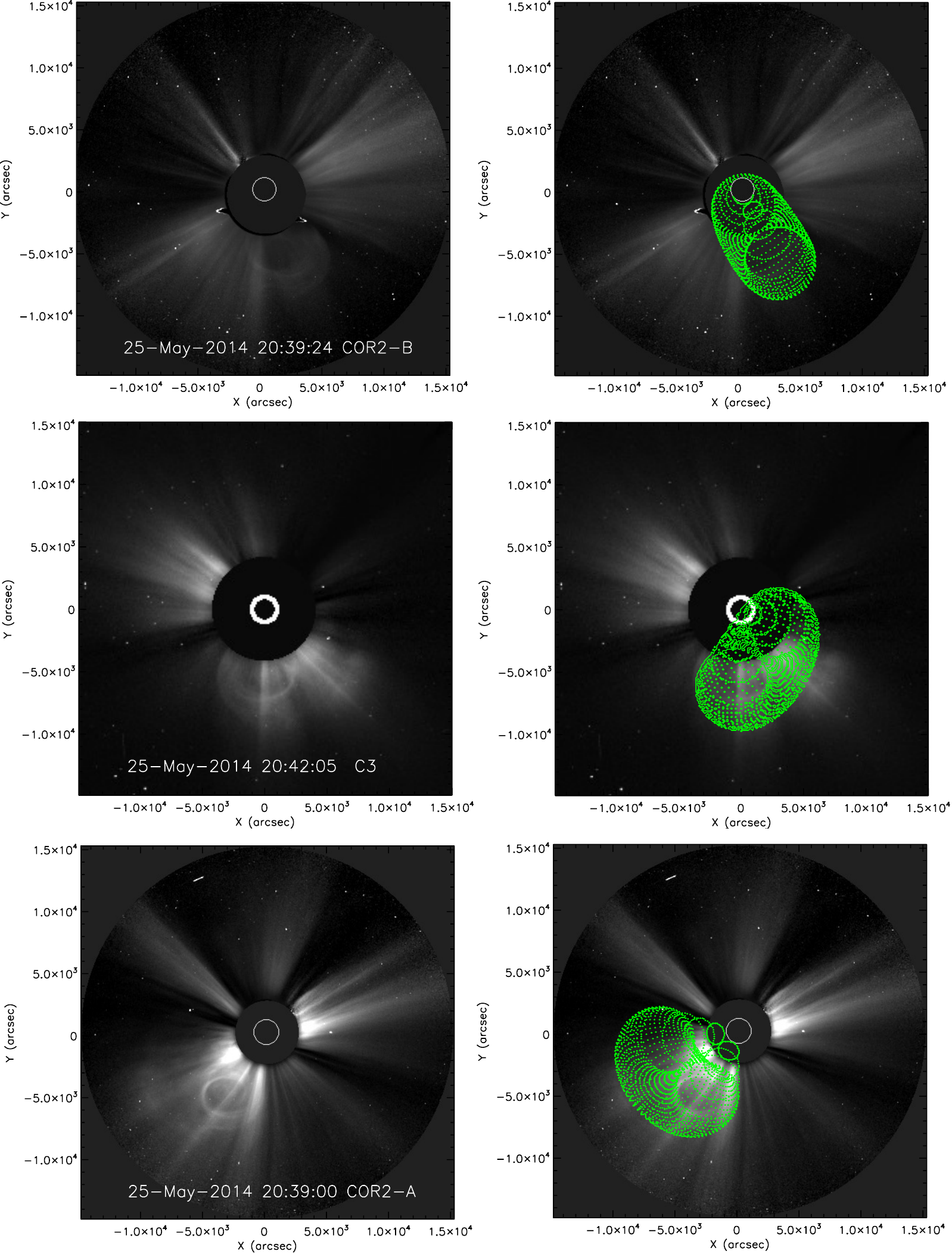}{0.45\textwidth}{(b)}
	      }
\caption{a) \textit{In situ} measurements of the magnetic cloud (1-minute averages) and the corresponding Forbush decrease of 2014 May 30. Top to bottom the panels show: 1) magnetic field strength (black) and fluctuations (gray); 2) magnetic field components (in Geocentric Solar Magnetospheric system); 3) plasma density (blue) and temperature (red); 4) plasma speed (black) and plasma-beta (gray); 5) relative hourly cosmic ray count by \textit{SOHO}/EPHIN.\\
b) GCS reconstruction for 2014 May 25 CME at 20:39 using coronagraphic images from \textit{STEREO-B}/COR2 (top), \textit{SOHO}/C3 (middle), and \textit{STEREO-A}/COR2 (bottom). Best fit parameters are: $\mathrm{longitude\,(Stonyhurst)}=6^{\circ}$, $\mathrm{latitude}=-20^{\circ}$, $\mathrm{tilt}=57^{\circ}$, $\mathrm{height}=18.2$\Rsun, $\mathrm{aspect\,ratio}=0.25$, and $\mathrm{half\,angle}=15^{\circ}$.
\label{fig5}}
\end{figure}

We select a near-Earth event of 2014 May 30, where an ejecta-only FD was recorded in the interplanetary space by \textit{SOHO}/EPHIN F-detector, which is suitable to detect ejecta-only FDs \citep[see \eg][]{heber15}. The event was associated with a magnetic cloud (MC) observed in the \insitu measurements of the \textit{Magnetic Field Instrument} \citep[MFI,][]{lepping95} and \textit{Solar Wind Experiment} \citep[SWE,][]{ogilvie95} onboard \textit{Wind} spacecraft (Figure \ref{fig5}a). The start/end of the MC is determined based on the drop in temperature and increase of the magnetic field strength, resulting in depressed plasma beta. In addition, throughout thus selected MC borders, signatures of the magnetic field rotation and smoothness are observed (first two panels of Figure \ref{fig5}a). The measured relative FD amplitude is $A=(3.3\pm0.1)\%$, where $A$ is the difference between FD minimum and ``quiet" time value derived as an average of the ``quiet" time measurements in a time period 1.5 days prior to the decrease. The standard error of 0.1 is also estimated based on the ``quiet" time \textit{SOHO}/EPHIN measurements.

In order to obtain CME initial parameters to run the FD model described in Section \ref{model2} (hereafter \textit{ForbMod}), the ICME recorded \insitu needs to be associated to a CME observed remotely by coronagraphs. We make a rough estimate of the possible ICME transit time ($TT=5$ days) based on the plasma speed at the leading edge ($v\approx350$\kmps) assuming constant speed and search for CMEs within a time window of roughly $\pm1$ day. The CME should be Earth-directed, therefore, we expect to see a halo or partial halo CME in the \textit{Large Angle Spectroscopic Coronagraph} \citep[LASCO,][]{brueckner95} or a CME with low-coronal signatures relatively close to the center of the solar disc, as seen in the EUV imagers. Using the \textit{SOHO}/LASCO CME Catalog\footnote{\url{https: //cdaw.gsfc.nasa.gov/CME list/}} we identified the most likely CME, a very slow partial halo first detected 2014 May 25 at 10:00 UT by LASCO/C2 moving throughout LASCO/C3 field of view until early May 26. We do not observe any obvious on-disc low-coronal signatures in \textit{Atmospheric Imaging Assembly} \citep[AIA,][]{lemen12} EUV imagers onboard \textit{Solar Dynamics Observatory} (SDO) or in EUV imagers onboard \textit{Solar-Terrestrial Relations Observatory} spacecraft \citep[STEREO,][]{kaiser08}. \textit{STEREO}/COR1 CME Catalog\footnote{\url{https://cor1.gsfc.nasa.gov/catalog/}} reports a faint flux-rope type eruption at 08:00 UT moving south in \textit{STEREO-B} and south-east in \textit{STEREO-A}.

In order to determine the CME initial parameters we use the Graduated cylindrical shell (GCS) model \citep{thernisien06,thernisien09,thernisien11} which is used to reconstruct the FR structure of the CME using coronagraphic images from different vantage points. The FR is modelled as a self-similarly expanding hollow croissant with origin in the center of the Sun, conical legs, circular cross section and pseudo-circular front. The reconstruction is done by visual comparison of the coronagraphic images from three different vantage points and the modelled FR. At 20:39 UT the CME is seen in \textit{STEREO-A}/COR2, \textit{STEREO-B}/COR2, and \textit{SOHO}/C3, where especially prominent circular-shaped substructure is seen in all three spacecraft (Figure \ref{fig5}b). This prominent feature is thus used as an additional constraint to the GCS (along with the shape of the front), compensating for the fact that the origin of the CME is unknown (no low-coronal on-disc signatures). The best fit is obtained by changing three geometric and three positional parameters. The positional parameters are latitude, longitude and tilt, whereas geometry parameters are the aspect ratio (parameter related to the varying radius of the cross section across the croissant axis), the half-angle (the angle between the axis of the leg and the face-on axis of the croissant) and the height. In Figure \ref{fig5}b the best fit of the GCS reconstruction is shown, where it can be seen that the CME is directed south of the equatorial plane and slightly to the west with respect to the Sun-Earth line. It is heavily tilted and therefore in the equatorial plane has a relatively small cross-section and width (small aspect ratio and half angle).

\begin{figure*}
\gridline{\fig{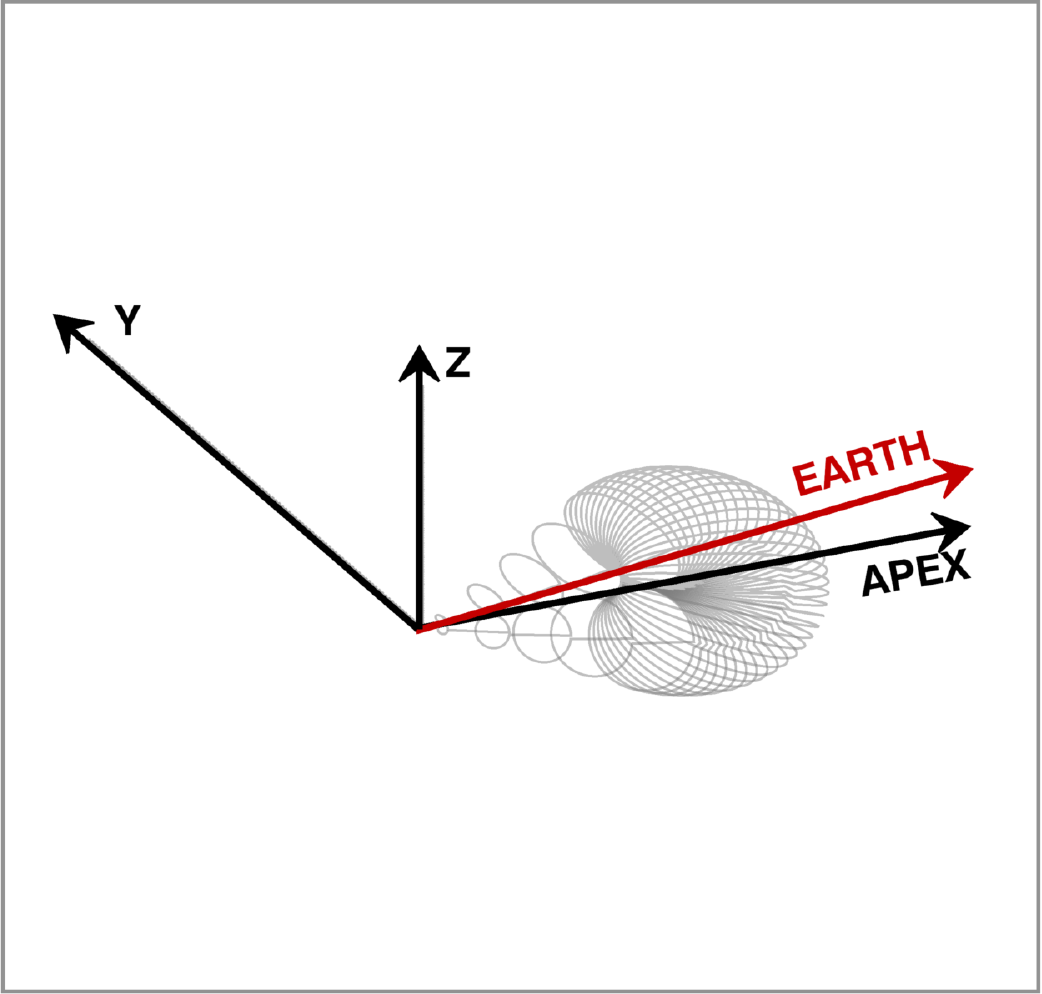}{0.43\textwidth}{(a)}
          \fig{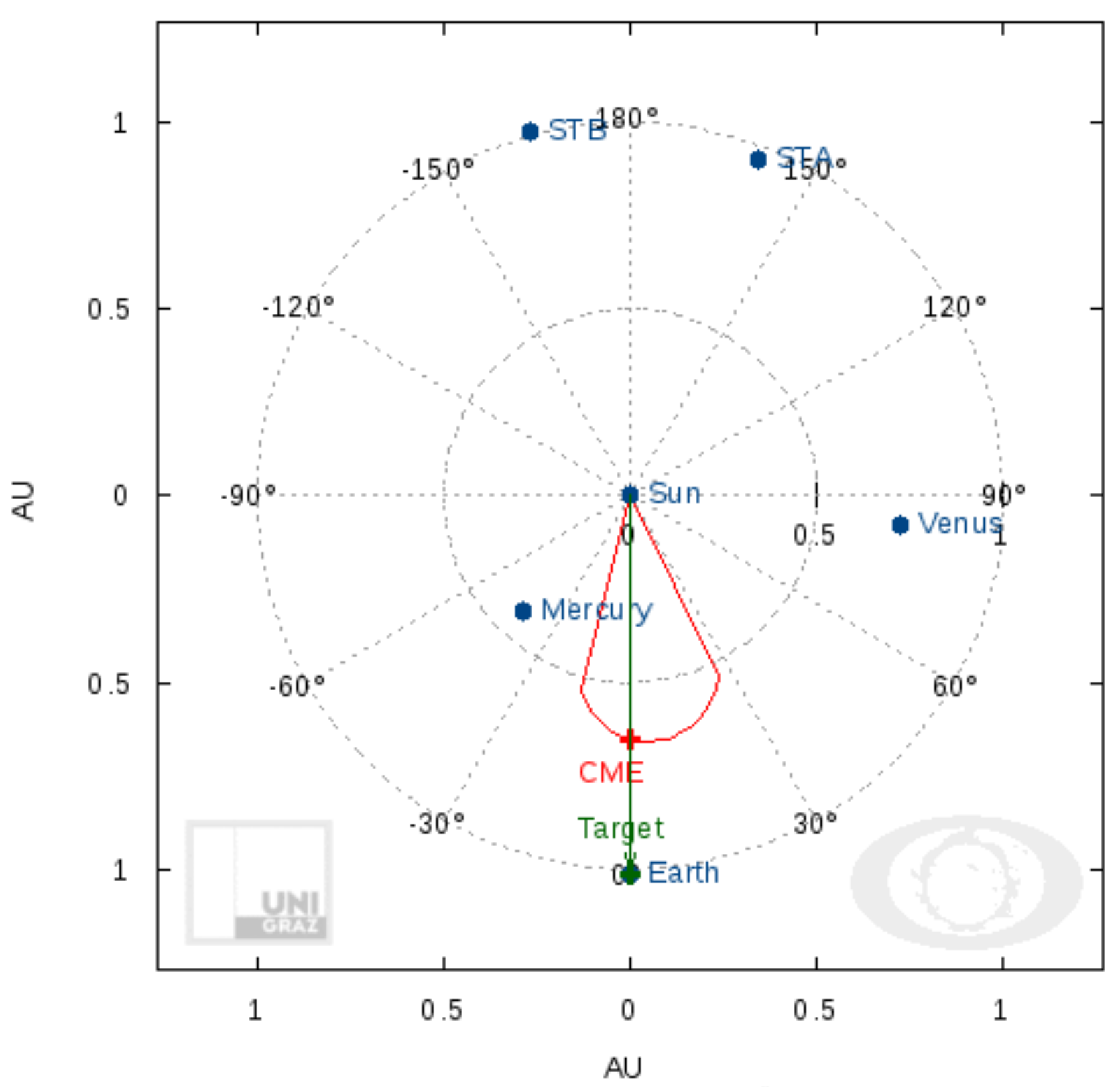}{0.43\textwidth}{(b)}     
          }
\caption{a) 3D plot of the reconstructed flux rope and the relative directions of the CME apex (black) and Earth (red) in the Heliocentric Earth Equatorial (HEEQ) system (Earth direction is along x-axis); b) CME geometry plot in the equatorial plane based on DBM (taken from \url{http://swe.ssa.esa.int}).
\label{fig6}}
\end{figure*}

Based on the GCS reconstruction we derive that the CME is Earth-directed with the apex slightly deflected southwest from the Sun-Earth line (Figure \ref{fig6}a). In order to test our CME-ICME association we use the Drag-Based Model \citep[DBM;][]{vrsnak13} for heliospheric propagation of ICMEs, which uses 2D CME cone geometry where the leading edge is initially a semicircle, spanning over the full angular width of the CME and flattens as it evolves in time \citep{zic15,dumbovic18}. As an input we use CME properties obtained by GCS. To obtain the initial kinematical properties of the CME, GCS reconstruction is performed on a series of coronagraph images 2014 May 25, 20:08 -- 22:24~UT, where the DBM initial height, time and velocity is given by the last GCS kinematical measurement ($R_0=21.9$\Rsun\, at 22:24, $v_0=355$\kmps). The width of the cone in the equatorial plane was estimated using $\omega_{\mathrm{max}} - (\omega_{\mathrm{max}} - \omega_{\mathrm{min}})\times\mathrm{tilt}/90$, where $\omega_{\mathrm{max}}$ and $\omega_{\mathrm{min}}$ are the maximum and minimum possible widths \citep[face-on and edge-on widths according to][]{thernisien11}, and the tilt is the angle of the croissant axis with respect to the equatorial plane. The estimated CME width in the equatorial plane is found to be $ \omega=20^{\circ}$. Based on the \insitu measured plasma speed (Figure \ref{fig5}a) we estimate that the ambient solar wind speed is $w\approx330$\kmps and we treat the drag parameter $\gamma$ as a free parameter of the DBM to obtain the most likely arrival time and speed roughly in agreement with the observed arrival time 2014 May 30 12:00 UT and arrival speed $v=350$\kmps. The best fit is obtained for $\gamma\approx0.5$\gammaunit. This is a quite high value and indicates a relatively dense and slow solar wind, and low CME mass \citep[see \eg][for details]{vrsnak13}, roughly in agreement with \insitu and white--light observation of solar wind and CME, respectively.

\begin{figure*}
\gridline{\fig{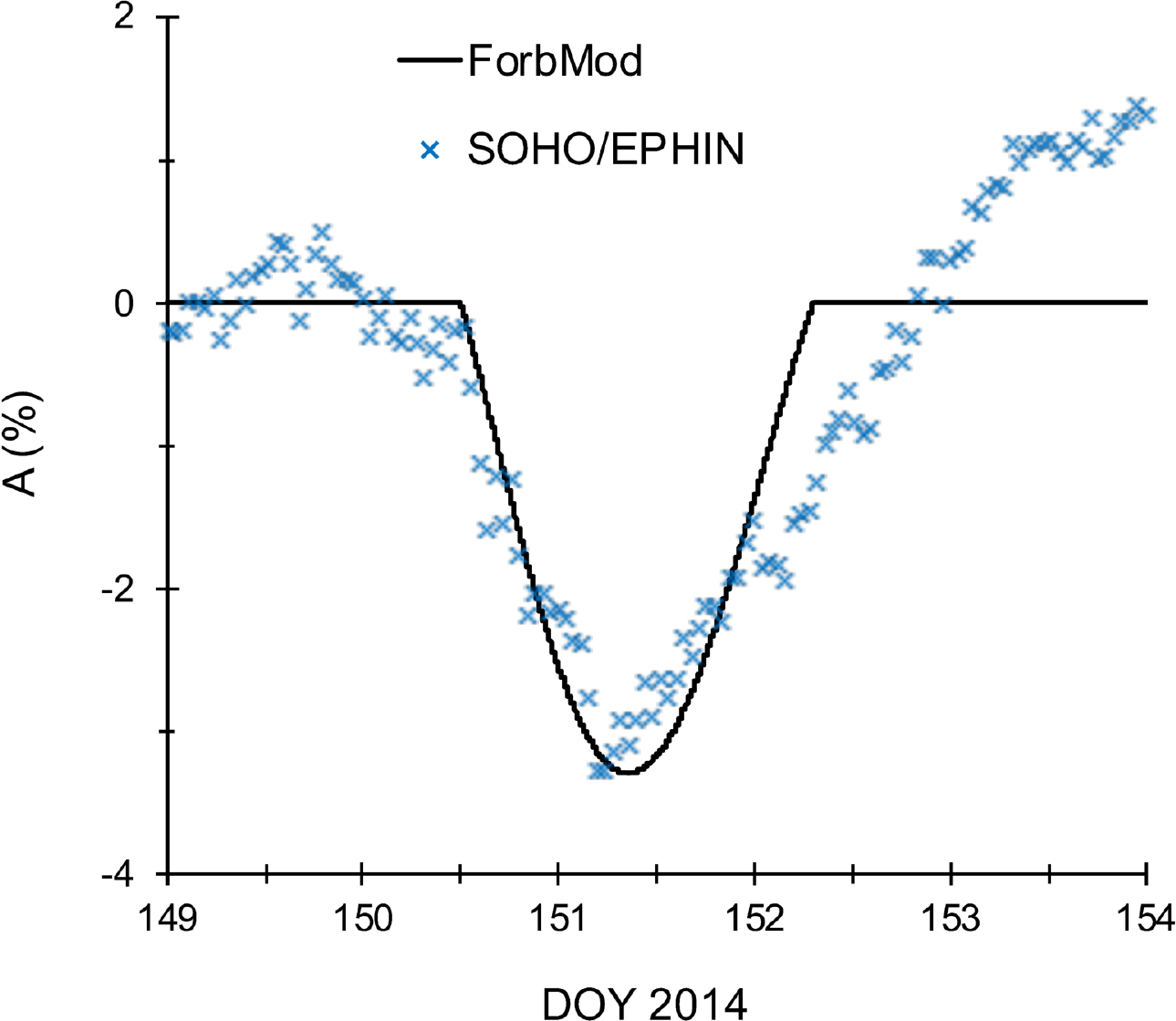}{0.43\textwidth}{(a)}
          \fig{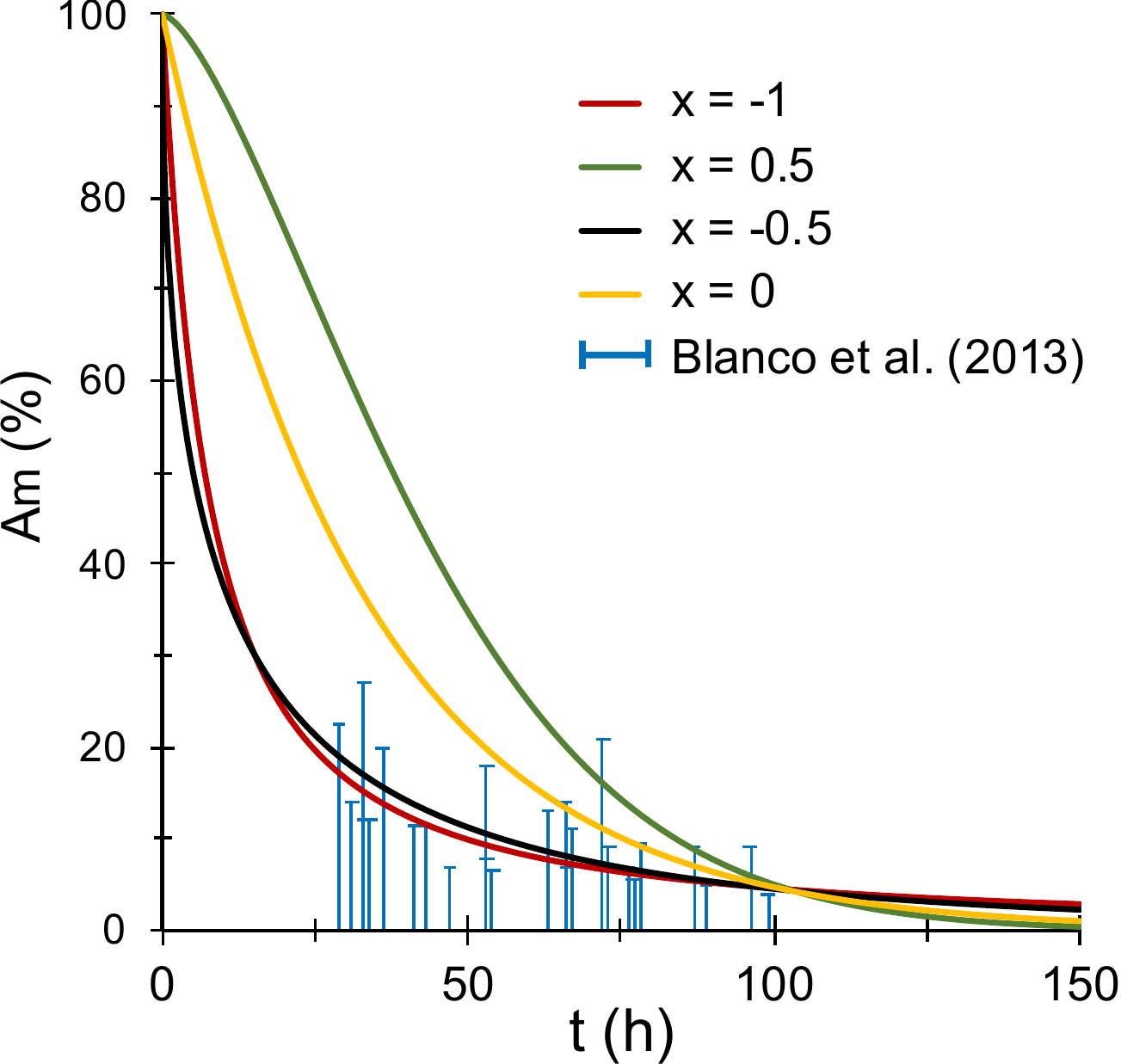}{0.43\textwidth}{(b)}   
          }
\caption{a) \textit{ForbMod} results (black line) for the ``best fit" value of the diffusion coefficient near Earth ($D_E=1.22$, 0.55, 0.68, and 0.45 in $10^{19}$\Dunit for $x=0$, 0.5, -0.5, and -1, respectively) against FD observation with \textit{SOHO}/EPHIN (blue crosses) for the event 2014 May 30. FD profile is converted from radial dependence into time dependence based on the ICME speed. The start and end of \textit{ForbMod} FD (black line) correspond to the shaded area in Figure \ref{fig5}a.\newline
b) \textit{ForbMod} time evolution of the FD magnitude for the event 2014 May 30 for different types of expansion \textit{vs.} observation \citep[data points presented by vertical lines are taken from Figure 8 in][for explanation see main text]{blanco13b}.
\label{fig7}}
\end{figure*}

Next we use GCS results to obtain the FR initial radius. For that purpose we calculate the radius of the croissant cross-section in the direction of the Earth, using formalism described by \citet{thernisien11} (the direction of Earth in the croissant face-on system was found to be 16$^{\circ}$ from the apex). We obtain $a_0=3.5$\Rsun\, at the start distance of $R_0=18.2$\Rsun\, and the corresponding transit time from $R_0$ to Earth (\ie diffusion time) is $t=111.3~$h. We treat the diffusion coefficient at Earth as a free parameter and search for the ``best fit", where the depression in  the center of the FR (\ie the FD magnitude) is equal to the measured FD magnitude $A=(3.3\pm0.1)\%$. Since all expansion cases show qualitatively the same results (Figure \ref{fig3}a), the best fit curve to the observed FD magnitude is identical for all expansion cases, however, with different corresponding values of the diffusion coefficient near Earth, $D_E$ (Figure \ref{fig7}a). It can be seen in Figure \ref{fig7}a that the observed FD profile is generally described well by \textit{ForbMod}, however there is a slight deviation at the rear of the FR, which could indicate deviation from the circular cross section. Considering the fact that the \insitu measurements show an asymmetric magnetic field profile with a declining magnetic field strength (first two panels in Figure \ref{fig5}a), one could imagine a slightly deformed FR which should be in such a case somewhat squeezed at the frontal side of the disturbance.

The  calculated temporal variation of the FD magnitude depends on the expansion case (Figure \ref{fig7}b), due to different $D_E$, as is expected from the analysis presented in Figure \ref{fig3}. To check whether the time evolution of the FD magnitude for the 2014 May 30 CME is quantitatively realistic, we compare it to the results of the statistical study by \citet{blanco13b}. In their statistical study, \citet{blanco13b} used spacecraft measurements from a Saphire Cherenkov detector onboard \textit{Helios 1} and \textit{Helios 2}, that is mainly sensitive to $\sim1$ GV particles, comparable to the response of the \textit{SOHO}/EPHIN \citep{kuhl15}. \citet{blanco13b} related the decrease in the C-detector (FD magnitude) versus the time of flight of the CME for a sample of ICMEs detected at \textit{Helios 1} and \textit{Helios 2}. We can use this statistical study to estimate whether the expected time evolution is quantitatively realistic. It should be noted that \citet{blanco13b} did not measure the FR part of the FD amplitude separately from the shock/sheath part, but present only the total FD amplitude. Therefore, in Figure \ref{fig7} we represent their measurements with error bars, where the value zero corresponds to no contribution from the FR part to the total FD amplitude, whereas the highest value corresponds to cases where the total FD amplitude is caused solely by the FR. It can be seen in Figure \ref{fig7}b that the time-evolution curve goes through at least some part of those presented \citet{blanco13b} measurements. However, a more detailed statistical analysis is needed for a more conclusive quantitative analysis, using only the FR part of the total FD amplitude. Nevertheless, we would still expect a significant scatter due to different events with different expansion characteristics. The time evolution for a specific expansion type could be \eg\, tested on an event detected at two different times, \ie radially aligned spacecraft at different heliospheric distances.

\begin{figure}
\centerline{\includegraphics[width=0.6\textwidth]{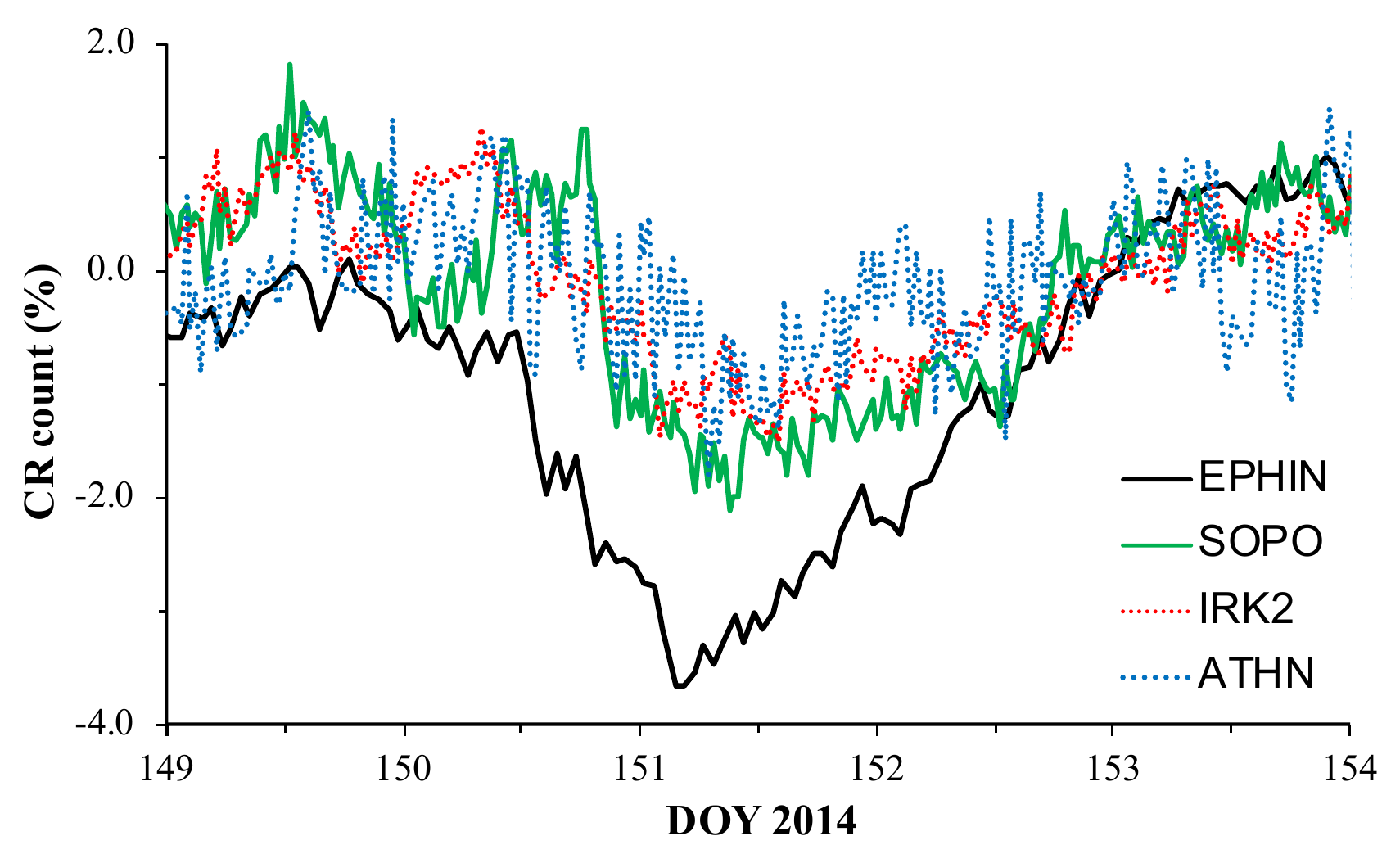}}
\caption{1-hour \textit{SOHO}/EPHIN and 30-minute neutron monitor relative cosmic ray counts during the 2014 May 30 FD. Neutron monitor measurements (provided by the Neutron Monitor Data Base, NMDB, \url{http://www.nmdb.eu/}) are corrected for pressure and efficiency, and correspond to different cutoff rigidity: SOPO=South Pole ($P=0.10$\,GV, $\mathrm{altitude}=2820$\,m), IRK2=Irkustk2 ($P=3.64$\,GV, $\mathrm{altitude}=2000$\,m), ATHN=Athens ($P=8.53$\,GV, $\mathrm{altitude}=260$\,m).}
\label{fig8}
\end{figure}

Finally, we estimate whether the values of $D_E$ are quantitatively realistic. We can see that $D_E$ is around $\sim10^{19}$\Dunit, roughly in agreement with \citet{cane95} and values obtained from the typical empirical expression used in numerical models \citep{potgieter13}. Furthermore, we determine the limit value of the $D_E$ after which FD magnitude at Earth is 0 for all expansion cases, and we find that the critical value is $D_{L}\approx3\times10^{19}$\Dunit. According to Equations 23 and 24 by \citet{potgieter13} for a magnetic field of $B\approx12$ nT this would roughly correspond to particle rigidity of $P_L\approx3$ GV, where $P_L$ can be regarded as an upper limit value of the rigidity where an FD is expected to be observed. Therefore, neutron monitors of different rigidity cutoffs offer convenient means to test this estimation. In Figure \ref{fig8} it can be seen that a small effect is observed in all selected neutron monitors, although the effect is almost indistinguishable due to daily variations, which are typically of the order of $\approx1\%$ \citep{parker64,tiwari12}. We note that although SOPO has a very low geomagnetic cutoff ($P=0.1$\,GV), we rather expect it to respond to particles $P>1$\,GV, due to atmospheric cutoff \citep[see \eg][and references therein]{clem00,moraal00}.  IRK2 and ATHN respond roughly to particles of $P>3.5$\,GV and $P>8.53$\,GV, respectively, where no effect is expected, but a small effect is however observed. This could be related to the fact that adiabatic cooling was not taken into account. On the other hand, it might also indicate that the rigidity dependence of the perpendicular diffusion coefficient might be different to the parallel one, as opposed to the \textit{ad hoc} assumption that $D_{\perp}$ scales with $D_{\parallel}$, used in our calculation of $P_L$, based on Equations 23 and 24 by \citet{potgieter13}. In either case, a reliable conclusion would require a more comprehensive analysis which is beyond the scope of this study. 

\section{Conclusion}
\label{conclusion}

The presented analytical diffusion--expansion Forbush decrease (FD) model \textit{ForbMod} relies on well established CME observations such as enhanced magnetic field and expansion. The model qualitatively reproduces well the established FD observations of roughly symmetrical FD profile and amplitude progressively decreasing with CME transit time. Quantitatively the model is determined by a complex interplay of the diffusion and expansion depending on the initial conditions (CME radius and diffusion coefficient), as well as the CME expansion type, \ie competing between the decrease of the magnetic field strength and the increase of the cross sectional area.

Several simplifications were used in order to model FDs analytically. We assume constant CME velocity and self similar expansion (no change in the shape) which can be relatively easily justified for slower CMEs but might be an important factor for very fast CMEs. In addition, we assume that the GCR density outside of the FR is constant, \ie that the change of the outside density of $\approx3\%$ throughout the evolution of the FR to 1 AU will not notably influence the filling rate of GCRs into the FR. We provide an estimate which justifies this assumption and furthermore, with the same estimates we can justify the applicability of the model to simulate the FR part of two-step FDs. Finally, we assume that GCRs enter the FR only \textit{via} diffusion and neglect possible contributions from adiabatic cooling due to expansion or additional outflow/inflow due to magnetic reconnection. Nevertheless, despite these simplifications used, the model reproduces a number of FD observational properties and furthermore is able to explain the FD observed on 2014 May 30. The observed FD was fitted quite well by \textit{ForbMod} for all expansion types using only the diffusion coefficient as a free parameter, where the diffusion parameter was found to be in the expected range of values.

We conclude that in general the model is able to explain the global properties of FD caused by FR, although there are indications of possible small deviations due to local properties and/or simplifications. These should be the subject of further studies, \eg\, a statistical study using constrains on the expansion type by \insitu measurements and/or FR forward modelling. Further testing of the model should also include multi-spacecraft FD measurements. Nevertheless, we note that even in the current form \textit{ForbMod} can be used to better understand the underlying physics in case studies.

\acknowledgments

The research leading to these results has received funding from the European Union's Horizon 2020 research and innovation programme under the Marie Sklodowska-Curie grant agreement No 745782. B. V. and M. D. acknowledge financial support by the Croatian Science Foundation under project 6212 ``Solar and Stellar Variability". M.T. acknowledges the support by the FFG/ASAP Programme under grant no. 859729 (SWAMI) and from the Austrian Science Fund (FWF): V195-N16.

\clearpage

\bibliographystyle{plainnat}
\bibliography{REFs}

\end{document}